\documentclass[%
preprint,
showpacs,preprintnumbers,
 amsmath,amssymb,
 aps,
pre,
]{revtex4-1}
\pdfoutput=1
\usepackage{graphicx}% Include figure files
\usepackage{dcolumn}% Align table columns on decimal point
\usepackage{bm}% bold math
\usepackage{multirow}

\usepackage[colorlinks,linkcolor=blue,citecolor=blue]{hyperref}
\usepackage{subfigure}
\usepackage{color}
\usepackage{url}
\usepackage{hyperref}

\usepackage{amsmath,amssymb}
\usepackage{subfigure}
\usepackage{caption}    % \ContinuedFloatºê°ü
\usepackage{color}
\usepackage{url}
\usepackage{lineno,hyperref}
\usepackage{graphicx}% Include figure files
\usepackage{dcolumn}% Align table columns on decimal point
\usepackage{bm}% bold math
\modulolinenumbers[5]

\begin{document}

%%\begin{frontmatter}



\title{An Implicit Discrete Unified Gas-Kinetic Scheme for Simulations of Steady Flow in All Flow Regimes}
%%\thanks{A footnote to the article title}%

\author{Dongxin Pan}
\email{chungou@mail.nwpu.edu.cn}
\author{Chengwen Zhong}%
\email{Corresponding author: zhongcw@nwpu.edu.cn}
\author{Congshan Zhuo}%
\email{zhuocs@nwpu.edu.cn}

\affiliation{National Key Laboratory of Science and Technology on Aerodynamic Design and Research, Northwestern Polytechnical University, Xi'an, Shaanxi 710072, China.}

\date{\today}% It is always \today, today,

\begin{abstract}
This paper presents an implicit method for the discrete unified gas-kinetic scheme (DUGKS) to speed up the simulations of the steady flows in all flow regimes. The DUGKS is a multi-scale scheme finite volume method (FVM) for all flow regimes because of its ability in recovering the Navier-Stokes solution in the continuum regime and the free transport mechanism in rarefied flow, which couples particle transport and collision in the flux evaluation at cell interfaces. In this paper the predicted iterations are constructed to update the macroscopic variables and the gas distribution functions in discrete microscopic velocity space. The lower-upper symmetric Gauss-Seidel (LU-SGS) factorization is applied to solve the implicit equations. The fast convergence of implicit discrete unified gas-kinetic scheme (IDUGKS) can be achieved through the adoption of a numerical time step with large CFL number. Some numerical test cases, including the Couette flow, the lid-driven cavity flows under different Knudsen number and the hypersonic flow in transition flow regime around a circular cylinder, have been performed to validate this proposed IDUGKS. The computational efficiency of the IDUGKS to simulate the steady flows in all flow regimes can be improved by one or two orders of magnitude in comparison with the explicit DUGKS.
\end{abstract}

\pacs{47.45.Ab, 02.70.-c, 47.11.Df}% PACS, the Physics and Astronomy

\keywords{implicit scheme \sep discrete unified gas-kinetic scheme \sep LU-SGS \sep all flow regimes}

\maketitle
% \linenumbers

\section{Introduction}
\label{Introduction}
%%\section{\label{sec:level1}Introduction}
%%\section{\label{sec:level1}Introduction}
The simulations for flows over a wide range of Knudsen numbers becomes a challenging issue for numerical modelling. Different flow physics in large variation of temporal and spatial scales cause a difficulty. In the case that the particle mean free path is of the same order as or even larger than representative physical length scale, gas can never be modeled as continuum \cite{sun2002direct}. The particle based methods may solve this kind of problem. But in continuum flow regime, its computational cost becomes unaffordable \cite{xu2010unified}.

To provide delicate numerical dissipation for particle collisions in flows especially for non-equilibrium phenomenon in shock-wave structure, the gas-kinetic scheme (GKS) is proposed by Xu et al \cite{prendergast1993numerical}. The GKS is able to give a more exact description for highly non-equilibrium flows than that of Navier-Stokes equation \cite{PhysRevE.95.053307}. Several studies have been done on implicit GKS for a better computation efficiency. Chit et al \cite{chit2004implicit} applied approximate factorization and alternating direction-implicit (AF-ADI) method on GKS. With a large Courant-Friedrichs-Levy (CFL) number, a fast convergence is achieved in simulation of the invicid compressible flows on structured grid. Compared to explicit GKS, this method even obtains a better accuracy. Li et al \cite{li2014implicit} proposed an implicit GKS based on matrix free Lower-Upper Symmetric Gauss Seidel (LU-SGS) time marching scheme for simulation of hypersonic inviscid flows on unstructured mesh. This method can be easily implemented on an hybrid unstructured mesh and the good robustness of this method is achieved. The issue is that the GKS is only valid for continuum flows.

Many kinetic schemes, such as the discrete ordinate method (DOM), can obtain accurate solutions in kinetic regimes but fail to simulate continuum flows efficiently because of the use of temporal and spatial scales on the order of particle collision time and mean free path \cite{xu2010unified}. Many other asymptotic preserving kinetic schemes successfully extend their validation to continuum invicid flows. But they still cannot capture mass and momentum transport in boundary layer \cite{dimarco2013asymptotic}. The unified gas kinetic scheme (UGKS) is a finite volume method (FVM) proposed to include all simulation scales from Navier-Stokes solution to kinetic regime. Based on the Boltzmann BGK model \cite{bhatnagar1954model}, it couples particle transport and collision process. In the reconstruction of the gas distribution function at the cell interface, the integral solution of the Boltzmann BGK model is applied. So, the numerical time step is not limited to the particle relaxation time \cite{xu2010unified}. In the update of the flow field, the UGKS has to compute the flux of macroscopic variables with moments for gas distribution function, which introduces additional computational cost in comparison with DOM. To reduce the computational cost and simplify computational process, the DUGKS for simulation of the flows with all Knudsen numbers is proposed \cite{guo2015discrete}. Different from the UGKS, the implicit treatment of the collision term is removed and the transformation of gas distribution function is employing with collision. In the flux evaluation, the gas distribution function is reconstructed at the cell interface along the characteristic line. So, the multi-scale dynamics in flows is described but the formulation of the numerical method is greatly simplified \cite{guo2013discrete}.

For UGKS, the matrix-free LU-SGS method for both continuum and rarified flow simulations has been constructed by Zhu et al (2016) \cite{zhu2016implicit}. In their work, both macroscopic and microscopic governing equations are implicitly coupled. To treat the collision term in an implicit way, the gas distribution function for equilibrium state is predicted by updating macroscopic variables implicitly. The governing equation for the gas distribution function is fully discretized in an implicit form. Finally, both of implicit governing equations of macroscopic variables and gas distribution function are solved iteratively based on the LU-SGS method. Compared to explicit UGKS, the implicit UGKS has a much faster convergence and the same accuracy in simulation for all flow regimes. But compared to DUGKS, the formulation of flux evaluation of UGKS is still too complicated. To simplify the computational process and improve the computational efficiency, it is necessary to develop an implicit method for DUGKS.

This paper is aimed to proposed an implicit method for DUGKS. In explicit DUGKS, the governing equation for macroscopic variables is not required. But in the IDUGKS, the BGK collision term should be treated in an implicit way. As a result, macroscopic variables should be updated implicitly in every implicit time step. Implicit discretization is applied in FVM governing equation for gas distribution function. In our work, the matrix free LU-SGS method is still used for discretizing and solving the linear systems derived from the implicit predicted treatment.

The rest of the paper is organized as follows. The classical DUGKS for all Knudsen numbers proposed by Guo et al \cite{guo2015discrete}, the matrix-free LU-SGS scheme and the implementation of IDUGKS are described in Section \ref{sec:IDUGKS}. Then, the numerical validations and the Couette flow in continuum flow regime, several lid-driven cavity flow cases under different Knudsen numbers, the hypersonic circular cylinder case are carried out to show the accuracy and the reliability of the present IDUGKS method in Section \ref{sec:NUMERICAL RESULTS}. Finally, some remarks concluded from this study are grouped in Section \ref{sec:conlustion}.


\section{Implicit method for discrete unified gas-kinetic scheme}\label{sec:IDUGKS}
\subsection{Discrete unified gas kinetic scheme}\label{sec:DISCRETE UNIFIED GAS KINETIC SCHEME}
In two dimensional problems, the DUGKS is based on the Boltzmann BGK model which is written as
\begin{equation}\label{Eq01}
\frac{{\partial f}}{{\partial t}} + u\frac{{\partial f}}{{\partial x}} + v\frac{{\partial f}}{{\partial y}} = \frac{{{f^{eq}} - f}}{\tau },
\end{equation}
where $f$  and $f^{eq}$  are the gas distribution functions, which are the functions of space $\left( {x, y} \right)$, particle velocity  $\left( {u, v} \right)$, time  $t$, and internal variable $\xi $. $\tau $ is the particle collision time. ${f^{eq}}$ is the Maxwell distribution function which has the following form
\begin{equation}\label{Eq02}
{f^{eq}} = \rho {\left( {\frac{\lambda }{\pi }} \right)^{\frac{{K + 2}}{2}}}{e^{ - \lambda \left( {{{\left( {u - U} \right)}^2} + {{\left( {v - V} \right)}^2} + {\xi ^2}} \right)}},
\end{equation}
where $K$ is the internal freedom degree with $K = 3$ for the 2D diatomic molecule gas flows. The variable $\lambda  = {m \mathord{\left/{\vphantom {m {\left( {2RT} \right)}}} \right.
 \kern-\nulldelimiterspace} {\left( {2RT} \right)}}$,  $m$ is the molecular mass, $R$ is the Boltzmann constant, and $T$ is the temperature.  $\rho $ is the density, $U$ and $V$ are the $x$ and $y$ components of the macroscopic velocity in 2D, respectively. Note that, for the gas system with $K$ freedom degree, the square of internal variable $\xi$ can be taken as
 \begin{equation}\label{Eq03}
{\xi ^2} = \xi _1^2 + \xi _2^2 +  \cdots  + \xi _K^2 .
\end{equation}

Conservative flow variables can be obtained by moments of gas distribution function with microscopic variables.
 \begin{equation}\label{Eq04}
\left( \begin{array}{c}
\rho \\
\rho U\\
\rho V\\
\rho E
\end{array} \right) = \int {\left( \begin{array}{c}
1\\
u\\
v\\
\frac{1}{2}\left( {{u^2} + {v^2} + {\xi ^2}} \right)
\end{array} \right)} fdudvd\xi ,
\end{equation}
where $\rho E = \frac{1}{2}\rho \left( {{U^2} + {V^2} + \frac{{K + 2}}{{2\lambda }}} \right)$  is the total energy. In this formulation $\lambda$ has relation $p = {\rho  \mathord{\left/
 {\vphantom {\rho  {2\lambda }}} \right. \kern-\nulldelimiterspace} {2\lambda }}$  with pressure $p$ and density $\rho $ \cite{xu2001gas}.
The evolution of the gas distribution functions has no relation with internal freedom. So, two reduced distribution functions are constructed by moments with variable
\begin{equation}\label{Eq05}
\begin{aligned}
G &= \int {fd\xi }, \\
H &= \int {{\xi ^2}fd\xi },
\end{aligned}
\end{equation}
where $G$ denotes particle density and $H$ reflects internal energy. The evolution equations for this two new distribution functions can be rewritten according to Eq.~(\ref{Eq01})
\begin{equation}\label{Eq06}
\begin{aligned}
\frac{{\partial G}}{{\partial t}} + u\frac{{\partial G}}{{\partial x}} + v\frac{{\partial G}}{{\partial y}} &= \frac{{{G^{eq}} - G}}{\tau }, \\
\frac{{\partial H}}{{\partial t}} + u\frac{{\partial H}}{{\partial x}} + v\frac{{\partial H}}{{\partial y}} &= \frac{{{H^{eq}} - H}}{\tau },
\end{aligned}
\end{equation}
where ${G^{eq}}$ and ${H^{eq}}$ are reduced equilibrium distribution function for $G$ and $H$.  The forms of governing equations for $G$ and $H$ are the same. So they have the same way in evolution. For simplicity, we use a variable $\phi $ to represent them. Eq.~(\ref{Eq06}) can be rewritten as
\begin{equation}\label{Eq07}
\frac{{\partial \phi }}{{\partial t}} + u\frac{{\partial \phi }}{{\partial x}} + v\frac{{\partial \phi }}{{\partial y}} = \vartheta ,
\end{equation}
where $\vartheta $  represents the collision term. In the finite volume, Eq.~(\ref{Eq07}) can be integrated over a time step, in which the midpoint rule is used for the time integration of the convection term and the trapezoidal rule for the collision term. The discrete form of Eq.~(\ref{Eq07}) can be written as
\begin{equation}\label{Eq08}
\left( {{\phi ^{n + 1}} - \frac{{\Delta t}}{2}{\vartheta ^{n + 1}}} \right) = \left( {{\phi ^n} + \frac{{\Delta t}}{2}{\vartheta ^n}} \right) - \frac{{\Delta t}}{\Omega }\sum\limits_{k = 1}^{Nf} {F_k^{n + \frac{1}{2}}} ,
\end{equation}
where $F_k^{n + \frac{1}{2}}$ denotes flux through cell interface $k$ and $Nf$ denotes the number of cell interface of a cell. $\Omega $ is the volume of the cell. The numerical flux reads
\begin{equation}\label{Eq09}
F = \int {{u_n}\phi dS}.
\end{equation}

In above equation, ${u_n}$ represents velocity propagate to cell interface. The term $\left( {{\phi ^{n + 1}} - \frac{{\Delta t}}{2}{\vartheta ^{n + 1}}} \right)$ and $\left( {{\phi ^n} + \frac{{\Delta t}}{2}{\vartheta ^n}} \right)$ are substituted by two new distribution functions $\tilde \phi $ and ${\tilde \phi ^ + }$ \cite{guo2015discrete}
\begin{equation}\label{Eq10}
\begin{aligned}
\tilde \phi  &= \frac{{2\tau  + \Delta t}}{{2\tau }}\phi  - \frac{{\Delta t}}{{2\tau }}{\phi ^{eq}}, \\
{{\tilde \phi }^ + } &= \frac{{2\tau  - \Delta t}}{{2\tau  + \Delta t}}\tilde \phi  + \frac{{2\Delta t}}{{2\tau  + \Delta t}}{\phi ^{eq}}.
\end{aligned}
\end{equation}

To obtain flux of the cell interface at $n + \frac{1}{2}$ time step, the distribution function at $n + \frac{1}{2}$ time step at the cell interface is required. In a half time step, Eq.~(\ref{Eq07}) is integrated and a relation is obtained \cite{guo2015discrete}
\begin{equation}\label{Eq11}
\bar \phi \left( {{{\mathord{\buildrel{\lower3pt\hbox{$\scriptscriptstyle\rightharpoonup$}}
\over x} }_{cf}}, {t_{n + \frac{1}{2}}}} \right) = {\bar \phi ^ + }\left( {\mathord{\buildrel{\lower3pt\hbox{$\scriptscriptstyle\rightharpoonup$}}
\over x}  - \mathord{\buildrel{\lower3pt\hbox{$\scriptscriptstyle\rightharpoonup$}}
\over u} \frac{{\Delta t}}{2}, {t_n}} \right),
\end{equation}
where $\bar \phi $ and ${\bar \phi ^ + }$ are two reduced distribution function which reads
\begin{equation}\label{Eq12}
\begin{aligned}
\bar \phi  &= \phi  - \frac{{\frac{{\Delta t}}{2}}}{2}\vartheta  ,\\
{{\bar \phi }^ + } &= \phi + \frac{{\frac{{\Delta t}}{2}}}{2}\vartheta = \frac{{2\tau  - \frac{{\Delta t}}{2}}}{{2\tau  + \frac{{\Delta t}}{2}}}\bar \phi  + \frac{{\Delta t}}{{2\tau  + \frac{{\Delta t}}{2}}}{\phi ^{eq}}.
\end{aligned}
\end{equation}

Now, $\bar \phi $ is the distribution function in Eq.~(\ref{Eq08}) at half time step. Thee distribution of equilibrium state can be obtained by macroscopic variables. They are computed via integration in microscopic velocity space with $\bar \phi $. The distribution function ${\bar \phi ^ + }\left( {\mathord{\buildrel{\lower3pt\hbox{$\scriptscriptstyle\rightharpoonup$}} \over x}  - \mathord{\buildrel{\lower3pt\hbox{$\scriptscriptstyle\rightharpoonup$}} \over u} \frac{{\Delta t}}{2}, {t_n}} \right)$ is obtained along characteristic line,
\begin{equation}\label{Eq13}
{\bar \phi ^ + }\left( {{{\mathord{\buildrel{\lower3pt\hbox{$\scriptscriptstyle\rightharpoonup$}}
\over x} }_{cf}} - \mathord{\buildrel{\lower3pt\hbox{$\scriptscriptstyle\rightharpoonup$}}
\over u} \frac{{\Delta t}}{2}, {t_n}} \right) = {\bar \phi ^ + }\left( {{{\mathord{\buildrel{\lower3pt\hbox{$\scriptscriptstyle\rightharpoonup$}}
\over x} }_{cf}}, {t_n}} \right) - u{\frac{{\Delta t}}{2}}\frac{{\partial {{\bar \phi }^ + }}}{{\partial x}} - v{\frac{{\Delta t}}{2}}\frac{{\partial {{\bar \phi }^ + }}}{{\partial y}}.
\end{equation}

In above equation, the spatial derivatives $\frac{{\partial {{\bar \phi }^ + }}}{{\partial x}}$ and  $\frac{{\partial {{\bar \phi }^ + }}}{{\partial y}}$, the value at the cell interface  ${\bar \phi ^ + }\left( {{{\mathord{\buildrel{\lower3pt\hbox{$\scriptscriptstyle\rightharpoonup$}}
\over x} }_{cf}}, {t_n}} \right)$ are obtained by the least square method. Taking the simple schematic grid shown in Fig.~\ref{fig:Fig01}, the cells in reconstruction are presented.

With the cells adjacent to the cell $({x_0}, {y_0})$ as well as values of them, the fitting formulation is applied,
\begin{equation}\label{Eq14}
{\bar \phi ^ + } = \bar \phi _0^ +  + \varphi \left[ {\frac{{\partial {{\bar \phi }^ + }}}{{\partial x}}\left( {x - {x_0}} \right) + \frac{{\partial {{\bar \phi }^ + }}}{{\partial y}}\left( {y - {y_0}} \right)} \right] .
\end{equation}

To determine the spatial derivative terms in above equation, the least-squares regression equations can be reconstructed as
\begin{equation}\label{Eq15}\tiny
\sum\limits_{j = 1}^N {\left[ \begin{array}{c}
{\left( {{x_j} - {x_0}} \right)^2}\\
\left( {{x_j} - {x_0}} \right)\left( {{y_j} - {y_0}} \right)
\end{array} \right.} \left. \begin{array}{c}
\left( {{x_j} - {x_0}} \right)\left( {{y_j} - {y_0}} \right)\\
{\left( {{y_j} - {y_0}} \right)^2}
\end{array} \right]\left[ \begin{array}{l}
\frac{{\partial {{\bar \phi }^ + }}}{{\partial x}}\\
\frac{{\partial {{\bar \phi }^ + }}}{{\partial y}}
\end{array} \right] = \sum\limits_{j = 1}^N {\left[ \begin{array}{l}
\left( {{x_j} - {x_0}} \right)\left( {\bar \phi _j^ +  - \bar \phi _0^ + } \right)\\
\left( {{y_j} - {y_0}} \right)\left( {\bar \phi _j^ +  - \bar \phi _0^ + } \right)
\end{array} \right]} ,
\end{equation}
where $N$ denotes the total number of cells adjacent to the cell. The $\varphi $ is the limiter. In incompressible flows, $\varphi  = 1$. In compressible flow simulation, the Vankatacrishnan limiter is applied \cite{venkatakrishnan1995convergence}.
\begin{equation}\label{Eq16}
\varphi  = \left\{
\begin{array}{*{20}{c}}
\theta \left( {\frac{{\max \left( {\bar \phi _j^ + } \right) - \bar \phi _0^ + }}{{\bar \phi _{cf}^ +  - \bar \phi _0^ + }}} \right),&\bar \phi _{cf}^ +  > \bar \phi _0^ + ,\\
\theta \left( {\frac{{\min \left( {\bar \phi _j^ + } \right) - \bar \phi _0^ + }}{{\bar \phi _{cf}^ +  - \bar \phi _0^ + }}} \right),&\bar \phi _{cf}^ +  < \bar \phi _0^ + ,\\
1&\bar \phi _{cf}^ +  = \bar \phi _0^ + ,
\end{array}
\right.
\end{equation}
where function $\theta $ can be calculated  by using the following formulas
\begin{equation}\label{Eq17}
\theta \left( x \right) = \frac{{{x^2} + 2x}}{{{x^2} + x + 2}} .
\end{equation}

In this paper, this limiter is applied in simulation for hypersonic rarified flow around circular cylinder. After ${\bar \phi ^{n + \frac{1}{2}}}$ at the cell interface is obtained, the original distribution function $\phi _{cf}^{n + \frac{1}{2}}$ can be obtained
\begin{equation}\label{Eq18}
\phi _{cf}^{n + \frac{1}{2}} = \frac{{2\tau }}{{2\tau  + \frac{{\Delta t}}{2}}}\bar \phi _{cf}^{n + \frac{1}{2}} + \frac{{\frac{{\Delta t}}{2}}}{{2\tau  + \frac{{\Delta t}}{2}}}\phi _{cf}^{eq,n + \frac{1}{2}} .
\end{equation}

At a cell interface $k$, the numerical flux can be obtained by
\begin{equation}\label{Eq19}
F_k^{n + \frac{1}{2}} = \int {{u_n}\phi _{cf}^{n + \frac{1}{2}}d{S_k}} .
\end{equation}

In gas kinetic theory, collision term in Boltzmann BGK model which contains only one single relaxation time leads to a fixed Prandtl number \cite{xu2001gas}. In this paper, BGK-Shakhov model is applied to overcome this limitation, the reduced distribution function ${G^{eq}}$ and ${H^{eq}}$ for equilibrium state are modified as \cite{shakhov1968generalization},
\begin{equation}\label{Eq20}\small
\begin{aligned}
G_{\Pr }^{eq} &= {G^{eq}} + \left( {1 - \Pr } \right)\frac{{\mathord{\buildrel{\lower3pt\hbox{$\scriptscriptstyle\rightharpoonup$}}
\over u}  \cdot \mathord{\buildrel{\lower3pt\hbox{$\scriptscriptstyle\rightharpoonup$}}
\over q} }}{{5pRT}}\left( {\frac{{{{\mathord{\buildrel{\lower3pt\hbox{$\scriptscriptstyle\rightharpoonup$}}
\over u} }^2}}}{{RT}} - \alpha  - 2} \right){G^{eq}},\\
H_{\Pr }^{eq} &= {H^{eq}} + \left( {1 - \Pr } \right)\frac{{\mathord{\buildrel{\lower3pt\hbox{$\scriptscriptstyle\rightharpoonup$}}
\over u}  \cdot \mathord{\buildrel{\lower3pt\hbox{$\scriptscriptstyle\rightharpoonup$}}
\over q} }}{{5pRT}}\left[ {\left( {\frac{{{{\mathord{\buildrel{\lower3pt\hbox{$\scriptscriptstyle\rightharpoonup$}}
\over u} }^2}}}{{RT}} - \alpha } \right)\left( {K + 3 - \alpha } \right) - 2K} \right]RT{G^{eq}} ,
\end{aligned}
\end{equation}
where $\alpha $ represents $\alpha $-dimensional problem.

\subsection{Matrix-free LU-SGS scheme}\label{sec:MATRIX-FREE LU-SGS SCHEME}
Now, the distribution function can be updated by FVM scheme
\begin{equation}\label{Eq21}
{\tilde \phi ^{n + 1}} = {\tilde \phi ^{ + ,n}} - \frac{{\Delta t}}{\Omega }\sum\limits_{k = 1}^{Nf} {F_k^{n + \frac{1}{2}}}.
\end{equation}

For simulation of steady state, implicit scheme can be constructed by using the backward Euler method at $n + 1$ time step
\begin{equation}\label{Eq22}
\frac{{{{\tilde \phi }^{n + 1}} - {{\tilde \phi }^n}}}{{\Delta t}}\Omega  = {M^n} + \left( {\frac{{\partial M}}{{\partial \tilde \phi }}} \right)\Delta \tilde \phi  + {Q^n} + \left( {\frac{{\partial Q}}{{\partial \tilde \phi }}} \right)\Delta \tilde \phi ,
\end{equation}
where $M$ denotes integration of flux around all interfaces and $Q$ denotes source term, they can be written as
\begin{equation}\label{Eq23}
\begin{aligned}
M &=  - \sum\limits_{k = 1}^{Nf} {F_k^{n + \frac{1}{2}}} ,\\
Q &=  - \frac{{2\Delta t}}{{2\tau  + \Delta t}}\tilde \phi  + \frac{{2\Delta t}}{{2\tau  + \Delta t}}{\phi ^{eq}}.
\end{aligned}
\end{equation}

The $M$ in derivative term $\frac{{\partial M}}{{\partial \tilde \phi }}$ can be rewritten in delta form
\begin{equation}\label{Eq24}
M =  - \sum\limits_{k = 1}^{Nf} {{u_n}{S_k}\Delta \tilde \phi _k^{n + 1}}.
\end{equation}

The final convergence solution will not be affected by different algorithms for implicit fluxes. As a result, the distribution function at the cell interface is constructed by upwind scheme \cite{thomas1990implicit}. For cell i,
\begin{equation}\label{Eq25}
\Delta \tilde \phi _k^{n + 1} = \frac{1}{2}\left( {\Delta \tilde \phi _i^{n + 1} + \Delta \tilde \phi _k^{n + 1}} \right) + \frac{1}{2}sign\left( {\mathord{\buildrel{\lower3pt\hbox{$\scriptscriptstyle\rightharpoonup$}}
\over u}  \cdot {{\mathord{\buildrel{\lower3pt\hbox{$\scriptscriptstyle\rightharpoonup$}}
\over S} }_k}} \right)\left( {\Delta \tilde \phi _i^{n + 1} - \Delta \tilde \phi _k^{n + 1}} \right).
\end{equation}

In cell $i$, the final implicit equation is written as
\begin{equation}\label{Eq26}
\left( {\frac{{{\Omega _i}}}{{\Delta {t_{imp}}}} - \frac{{\partial M}}{{\partial {{\tilde \phi }_i}}} - \frac{{\partial {Q}}}{{\partial {{\tilde \phi }_i}}}} \right)\Delta {\tilde \phi _i} - \frac{{\partial M}}{{\partial {{\tilde \phi }_j}}}\Delta {\tilde \phi _j} = {M^n} + {Q^n}.
\end{equation}

After iterating over the whole flow field, a system of linear equations $\Delta \tilde \phi $ is obtained. This equations can be discretized and solved by using LU-SGS method \cite{yoon1988lower}. Different from explicit DUGKS, macroscopic variables and microscopic distribution function should be marching at the same time step so implicit predicted algorithm is applied to governing equation for macroscopic variables. For DUGKS, the scheme for updating of macroscopic variables reads

\begin{equation}\label{Eq27}
\frac{{{W^{n + 1}} - {W^n}}}{{\Delta t}}\Omega  =  - \sum\limits_{k = 1}^{Nf} {\int {\psi F_k^{n + \frac{1}{2}}dudv} },
\end{equation}
where $\psi  = {\left( {1, u, v, \frac{1}{2}\left( {{u^2} + {v^2} + {\xi ^2}} \right)} \right)^T}$ is microscopic variables. When updating macroscopic variables the numerical flux at half time step is still used. Eq.~(\ref{Eq27}) can be rewritten as implicit form
\begin{equation}\label{Eq28}
\frac{{{W^{n + 1}} - {W^n}}}{{\Delta t}}\Omega  = M_W^n + \left( {\frac{{\partial {M_W}}}{{\partial W}}} \right)\Delta W ,
\end{equation}
where the term $\frac{{\partial {M_W}}}{{\partial W}}$ is a Jacobian matrix. In this paper, the form in first-order Roe¡¯s scheme \cite{roe1981approximate} in which the Roe¡¯s flux can be linearized is applied in discretization for macroscopic predicted algorithm \cite{chen2000fast}. According to the idea of LU-SGS method, we first split this Jacobian matrix into three parts: the lower triangular matrix, the upper triangular matrix and the diagonal terms. Eq.~(\ref{Eq28}) can be rewritten as
\begin{equation}\label{Eq29}
\left( {L + U + D} \right)\Delta W = M_W^n ,
\end{equation}
with
\begin{equation}\label{Eq30}
\left\{ \begin{array}{l}
L = \sum\limits_{k \in L\left( i \right)} {\left( {\frac{1}{2}\frac{{\partial F\left( W \right)}}{{\partial W}} \cdot S - {{\left( {{\Lambda _c}} \right)}_{ik}}} \right)} ,\\
U = \sum\limits_{k \in U\left( i \right)} {\left( {\frac{1}{2}\frac{{\partial F\left( W \right)}}{{\partial W}} \cdot S - {{\left( {{\Lambda _c}} \right)}_{ik}}} \right)} ,\\
D = \left( {\frac{{{\Omega _i}}}{{\Delta {t_{imp}}}} + \sum\limits_{k \in Nf} {{{\left( {{\Lambda _c}} \right)}_{ij}}} } \right)I ,
\end{array} \right.
\end{equation}
where $L\left( i \right)$ represents $k < i$, and $U\left( i \right)$ represents $k > i$. ${\Lambda _c}$ has the form \cite{yoon1988lower}
\begin{equation}\label{Eq31}
{\left( {{\Lambda _c}} \right)_i} = \sum\limits_{k = 1}^{Nf} {\left( {\left| {{{\mathord{\buildrel{\lower3pt\hbox{$\scriptscriptstyle\rightharpoonup$}}
\over V} }_i} \cdot {{\mathord{\buildrel{\lower3pt\hbox{$\scriptscriptstyle\rightharpoonup$}}
\over n} }_k}} \right| + {C_i}} \right){S_k}},
\end{equation}
where ${\mathord{\buildrel{\lower3pt\hbox{$\scriptscriptstyle\rightharpoonup$}}\over V} _i}$ represents the macroscopic velocity vector in cell $i$. ${\mathord{\buildrel{\lower3pt\hbox{$\scriptscriptstyle\rightharpoonup$}} \over n} _k}$ represents unit normal vector of the interface with area ${S_k}$. Eq.~(\ref{Eq29}) is discretized and solved by using LU-SGS method. Finally, the distribution function and macroscopic variables are updated in the same implicit time step.
\begin{equation}\label{Eq32}
\begin{aligned}
{{\tilde \phi }^{n + 1}} &= {{\tilde \phi }^n} + \Delta {{\tilde \phi }_{imp}},\\
{W^{n + 1}} &= {W^n} + \Delta {W_{imp}} .
\end{aligned}
\end{equation}

In above procedure, the inner variable $\xi $ is continuous. But for $\alpha $-dimensional velocities $u$, $v$, velocity space is discretized. In this paper, Gauss-Hermite quadrature formula \cite{shizgal1981gaussian} or Newton-Cotes formula \cite{forsythe1977computer} is applied in integration under discrete velocity space.

\subsection{Boundary condition}\label{sec:BOUNDARY CONDITION}
In the IDUGKS, ghost cells are employed along boundary of flow field. In evaluation of explicit flux, variables in ghost cells are directly derived from corresponding inner cells as Ref.\cite{xu2001gas}. In prediction step of implicit scheme, to improve convergence efficiency, a governing equation of boundary is included. Boundary condition for implicit iteration is implemented on linearized relation between inner and ghost cells \cite{aoki2009numerical}. For implicit flux term $M$ and variable $\Pi$ (distribution function and macroscopic variables) to be updated, the linearized relation reads
\begin{equation}\label{Eq33}
\Delta \Pi _{ghost}^{n + 1} - \left( {\frac{{\partial M}}{{\partial \Pi }}} \right)_{inner}^n\Delta \Pi _{inner}^{n + 1} = 0 .
\end{equation}

On the solid wall, it is not so easy for distribution function to applied Eq.~(\ref{Eq33}). In this paper, diffusive reflection boundary condition is applied \cite{meng2014diffuse}.
\begin{equation}\label{Eq34}
\Delta f_{ghost}^{n + 1} = \Delta {\rho _{ghost}}{\left( {\frac{{{\lambda _{wall}}}}{\pi }} \right)^{\frac{{K + 2}}{2}}}{e^{ - {\lambda _{wall}}\left[ {{u^2} + {v^2} + {\xi ^2}} \right]}},
\end{equation}
where
\begin{equation}\label{Eq35}
\Delta {\rho _{ghost}} = \frac{{ - \int {\frac{1}{2}\left( {{u_n} + \left| {{u_n}} \right|} \right)\Delta {f_{inner}}dudvd\xi } }}{{\int {\frac{1}{2}\left( {{u_n} - \left| {{u_n}} \right|} \right)} {{\left( {\frac{{{\lambda _{wall}}}}{\pi }} \right)}^{\frac{{K + 2}}{2}}}{e^{ - {\lambda _{wall}}\left[ {{u^2} + {v^2} + {\xi ^2}} \right]}}dudvd\xi }} .
\end{equation}

At the beginning of computation at $0$ time step, the forward sweep can begin with boundary condition $\Delta W_{ghost}^{n + 1} = 0$ for macroscopic variables and $\Delta f_{ghost}^{n + 1} = 0$ for microscopic distribution function. This treatment will not cause any negative defect to accuracy and convergence.

\section{Numerical results}\label{sec:NUMERICAL RESULTS}
The present IDUGKS will be validated by test cases in different flow regimes. First, the test case of Couette flow is carried out to validate that the IDUGKS is a second-order accurate numerical method.  Then, the test cases of Lid-driven cavity flow in different Knudsen numbers are conducted to demonstrate that the IDUGKS is capable of simulating flows in all flow regime. Finally, the hypersonic flow in transition flow regime is performed to show the IDUGKS can treat the flow with shock wave. After comparison results in every case, the computational cost will be presented with residual curve. In all test cases the implicit CFL number is chosen as ${10^3}$. Comparison of computational efficiency measured with wall clock time between implicit and explicit method will be given at the end of this session. Some remarks will also be given.

\subsection{Couette flow}\label{sec:Couette flow}

The Couette flow is driven by two parallel plates. the top plate is moving in constant velocity and the other is static. The Reynolds number is chosen as 100 and Mach number is $0.1\sqrt 3 $. In discrete velocity space, $9 \times 9$ discrete points are distributed uniformly in $\left[ { - 2.5, 2.5} \right] \times \left[ { - 2.5, 2.5} \right]$. The velocity distributions along y-direction are plotted in Fig.~\ref{fig:Fig02} on uniform meshes within $\left[ {0, 1} \right] \times \left[ {0, 1} \right]$ of $20 \times 20$, $40 \times 40$, $80 \times 80$, $160 \times 160$. It can be seen that the numerical results are in excellent agreement with the analytical ones.

To test the convergence order of the IDUGKS, the results from simulations on different meshes have been used to compute the L2 errors in velocity field along y-direction. The L2 error is defined by
\begin{equation}\label{Eq2001}
E(U) = \frac{{\sqrt{\sum_y {|U(y)-U_e(y)|^2}}}}{{\sqrt{\sum_y {U_e(y)^2}}}} .
\end{equation}
where $U_e$ is the analytical solution of Couette flow case. The error in L2 norm with respect to mesh size is plotted in Fig.~\ref{fig:Fig03}, which shows a nearly second-order accuracy of the implicit scheme.

\subsection{Lid-driven cavity flows}\label{sec:Lid-driven cavity flow cases}

The first case is incompressible lid-driven cavity flow at Reynolds number $1000$ and Mach number $0.1\sqrt 3 $. The uniform mesh with $257 \times 257$ mesh points within $\left[ {0, 1} \right] \times \left[ {0, 1} \right]$ is chosen as computational domain. For discrete velocity space, $9 \times 9$ discrete points are distributed uniformly in $\left[ { - 2.5, 2.5} \right] \times \left[ { - 2.5, 2.5} \right]$. The streamlines are plotted in Fig.~\ref{fig:Fig06} and comparison result for V velocity profile along line $Y = 0.5$ is presented in Fig.~\ref{fig:Fig07}.

The V velocity profile obtained using IDUGKS is compared with that from explicit DUGKS and U. Ghia et al \cite{ghia1982high}. In continuum regime, the IDUGKS can obtain the same accuracy as the explicit method. Both of them are in good agreement with the previous literature. For a better demonstration for accuracy of the IDUGKS in description of incompressible flow compared to other numerical scheme, a list including the vorticity, the stream function, and $x$, $y$ location of vortex center are grouped in Table~\ref{table:Table1}.

In Table~\ref{table:Table1}, besides results from \cite{ghia1982high}, all the flow properties in incompressible cavity flow are also compared to results obtained from the lattice Boltzmann method (LBM) in work of Zhuo et al \cite{zhuo2012filter} and Hou et al \cite{hou1995simulation}. Comparison show that the difference between present results and the benchmark results is less than $2\% $. With a much larger CFL number than explicit method, IDUGKS requires much fewer iteration step to reach convergence. The residual curves for comparison are plotted in Fig.~\ref{fig:Fig08}.

In this case, the explicit DUGKS requires $255000$ iteration step to reach a residual of $5 \times {10^{ - 8}}$. To obtain the same convergence criterion the IDUGKS only needs $2040$ iteration step.

Now, we shift flow regime to slip flow regime. The Knudsen number is set to be $0.075$. According to definition of Knudsen number
\begin{equation}\label{Eq36}
Kn = \frac{\bar l}{L_{ref}} ,
\end{equation}
where $\bar l$ denotes the mean free path of particles, in this case, ${L_{ref}}$ is the reference length. In the test cases of cavity flow the reference length is the side length of the cavity.

The uniform mesh with $60 \times 60$  mesh cells within $\left[ {0, 1} \right] \times \left[ {0, 1} \right]$ is chosen as computational domain. For discrete velocity space, $60 \times 60$ discrete points are distributed uniformly in $\left[ { - 2.5, 2.5} \right] \times \left[ { - 2.5, 2.5} \right]$. In this case, the Gauss-Hermite quadrature formula is chosen for the integration of distribution function under discrete velocity space. The horizontal velocity profile and vertical profile are given in Fig.~\ref{fig:Fig09}.

In Fig.~\ref{fig:Fig09}, the velocity profiles obtained from the IDUGKS are compared with UGKS and DSMC presented in Ref.~\cite{zhu2016implicit}. The IDUGKS is able to reach the same accuracy as UGKS and particle based direct simulation Monte Carlo (DSMC). In slip flow regime like this, there is slip velocity on solid wall as shown in Fig.~\ref{fig:Fig09}. This phenomenon is different from continuum flow in which fluid does not have any motion on solid wall. Existence of slip velocity on wall is consistence with results from DSMC. The residual curves for comparison between the implicit and the explicit DUGKS are shown in Fig.~\ref{fig:Fig10}.

The convergence criterion for residual is also set to be $5 \times 10^{-8}$. The explicit DUGKS requires $11900$ iteration steps to obtain convergence while just $230$ is needed for the IDUGKS. Even though predicted iteration will cost some time, the implicit method still improve convergence efficiency for about $27$ times.

Next, the lid-driven cavity flow under $Kn = 1.0$ is simulated. This test is to validate the IDUGKS in simulation for rarefied flows. In this paper, Knudsen number takes its effect in computation by relation with Reynolds number $Re$ and Mach number $Ma$ \cite{laurendeau2005statistical}
\begin{equation}\label{Eq37}
Kn = \frac{Ma}{Re} \sqrt{\frac{\gamma \pi}{2}} ,
\end{equation}
where $\gamma$ is ratio of specific heats. The computational domain is also set to be $\left[ {0, 1} \right] \times \left[ {0, 1} \right]$ and discretized with $61 \times 61$ mesh points. $60 \times 60$ discrete velocity points are distributed in $\left[ { - 2.5, 2.5} \right] \times \left[ { - 2.5, 2.5} \right]$ space. In DUGKS, in order to avoid divergence, the error from discrete integration is not allow to be greater than $1\%$. As a result, the Gauss-Hermite quadrature formula is applied. The velocity profiles in X and Y direction are plotted in Fig.~\ref{fig:Fig11} and compared with results from Ref. \cite{zhu2016implicit}.

Results from the IDUGKS fit quite well with that from DSMC. In rarefied flow simulation, the IDUGKS is still be able to capture physical details of gas flow. Compared to particle based Lagrangian method, The DUGKS not only does not cause extra computational cost to trace the trajectory of particle motion, but also it can obtain the same accuracy. The convergence history is presented with residual curve in Fig.~\ref{fig:Fig12}.

For the IDUGKS, $190$ iteration steps are required while the explicit method uses $9600$ iteration steps. In rarefied flow, a much higher efficiency than explicit method is still obtained by the IDUGKS. To verify the IDUGKS for all flow regimes, in final case of lid-driven cavity flow, test for $Kn = 1.0$ is simulated. The physical space of $\left[ {0, 1} \right] \times \left[ {0, 1} \right]$ discretized with $61 \times 61$ mesh points and microscopic velocity space of $\left[ { - 2.5, 2.5} \right] \times \left[ { - 2.5, 2.5} \right]$ with $60 \times 60$ discrete velocity points are chosen. In discrete integration, the Gauss-Hermite quadrature formula is used. Flow structure in this case where $\bar l \gg L_{ref}$ is described with horizontal and vertical velocity profile along the central line.

In kinetic regime, solution of the IDUGKS nicely fits with the one from DSMC and UGKS in Ref. \cite{zhu2016implicit}. Residual curves are plotted in Fig.~\ref{fig:Fig14}.

In all flow regimes, from continuum flow to kinetic regime, the DUGKS gives the same exact description for flow structure as UGKS and DSMC. But its formulation is much more simple and easy to be implemented. Implementation of implicit scheme greatly reduces iteration steps thus improves efficiency of the original explicit DUGKS. The IDUGKS is a reliable and efficient method for simulation of multi-scale flows.

\subsection{Hypersonic rarefied flow around circular cylinder}\label{sec:Hypersonic rarefied flow around circular cylinder}

In this case, supersonic flow passing through a circular cylinder with $Kn = 1.0$ and $Ma = 5.0$ is computed. The physical space is discretized with $4800$ mesh cells and microscopic velocity space of  $\left[ { - 15, 15} \right] \times \left[ { - 15, 15} \right]$ with $90 \times 90$ discrete velocity points are chosen. In discrete integration, the Gauss-Hermite quadrature formula is used. In reconstruction at the cell interface, the Vankatakrishnan limiter is implemented. The density contour and pressure contour are presented in Fig.~\ref{fig:Fig15a} and Fig.~\ref{fig:Fig15b}.

The pressure distribution on the surface of circular cylinder is extracted to compared with results in Ref. \cite{zhu2016implicit}. Because the flow is symmetric, we just show the part within ${180^o}$ from the leading edge to the trailing edge. Comparison results are presented in Fig.~\ref{fig:Fig16}.

Like lid-driven cavity flow, in rarefied regime, there is slip on solid wall. Velocity vectors near circular cylinder are plotted in Fig.~\ref{fig:Fig17}. The mach number contour is also presented in Fig.~\ref{fig:Fig18}.

Slip on surface of solid wall is demonstrated with Fig.~\ref{fig:Fig17}. To give a quantitative description, the shear stress curve on surface of circular cylinder from leading edge to trailing edge is plotted and compared to results in previous literature \cite{zhu2016implicit} in Fig.~\ref{fig:Fig19}.

Heat flux distribution on the surface of circular cylinder is extracted to compared with results in Ref. \cite{zhu2016implicit}. In this paper, BGK-Shakhov model is applied to capture heat transfer. From leading edge to trailing edge, the heat flux curve fits pretty well with previous literature as shown in Fig.~\ref{fig:Fig20}.

Since about ${50^o}$ clockwise from leading edge, the shear stress begins to drop. This result is the same as previous work based on UGKS and DSMC. Residual curves of explicit and implicit DUGKS are plotted in Fig.~\ref{fig:Fig21} for comparison.

For a suitable growth rate of mesh around shock wave, the adaptive mesh refinement (AMR) is applied in this
case. The initial mesh with $2400$  mesh cells is shown in Fig.~\ref{fig:Fig22}. Mesh around shock wave  is refined using gradient and vorticity criteria. The final hybrid mesh with 4363 cells and pressure contour captured by it are presented in Fig.~\ref{fig:Fig23} and Fig.~\ref{fig:Fig24}.

In order to further explore the effects of unstructured hybrid mesh on accuracy and convergence, computation results and residual curve under refined mesh are compared with under structured mesh in this paper. Comparisons for pressure, shear stress and heat flux distribution on the surface of circular cylinder on different meshes are presented in Fig.~\ref{fig:Fig25}, Fig.~\ref{fig:Fig26} and Fig.~\ref{fig:Fig27}. To obtain the same accuracy, the number of cells used in refined mesh is fewer than structured grid. For the present two simulations with different kinds of meshes, the residual curves are plotted in Fig.\ref{fig:Fig28}. It can be observed that, the two simulations with structured and unstructured refined meshes reach the steady state quickly and only take 2040 and 1750 iteration steps respectively.

In all cases, the implicit DUGKS requires much fewer iteration steps than explicit DUGKS. To measure quantitatively the computational efficiency of numerical scheme proposed in this paper, comparison for computational cost between the IDUGKS and the explicit DUGKS is shown in Table~\ref{table:Table2} by wall clock time.

For cavity flow, acceleration rates are $89.9$, $27.4$, $26.8$, $28.5$ respectively. In hypersonic circular cylinder case, the acceleration rate becomes $30.7$.

In all, the present IDUGKS solver is applicable for flows from continuum regime to kinetic regime, from incompressible to hypersonic flows. Compared to explicit DUGKS, much lower computation time is required. The IDUGKS proposed in this paper shows excellent accuracy and efficiency.

\section{Conclusions}\label{sec:conlustion}
In this paper, an implicit DUGKS is constructed for all Knudsen number flows. The physics represented in DUGKS depends on coupling of particle transport and collision in flux evaluation at cell interface. DUGKS can obtain accurate solution for multi-scale flow problems. Since the application of variation of ratio between explicit time step and particle collision time makes DUGKS an asymptotic preserving method, a physical and a numerical time steps are used in the IDUGKS. In this paper, the coupled implicit macroscopic and microscopic iterative equations are applied, which increases the computation efficiency. Two iterative methods, LU-SGS method and point relaxation scheme, are implemented in prediction step. Comparison results for computational cost with explicit DUGKS proves a much better efficiency of the IDUGKS for steady state solution. In test cases, the scheme proposed in this paper is proved to have the same accuracy as DSMC. The present IDUGKS solver can be easily extended to the 3D case. Many technique for improvement, such as parallelization, immersed boundary (IB) method, can be developed. The IDUGKS constructed in this work is promising in its extensive applications.

\section*{Acknowledgements}
The work has been supported by the National Natural Science Foundation of China (Grant No. 11472219), the 111 Project of China (B17037), as well as the ATCFD Project (2015-F-016).

%% References

%%\bibliographystyle{elsarticle-num}
%%\bibliographystyle{model3-num-names}
\bibliography{IDUGKS}
%% not want to use model3-num-names.bst.

%%\newpage %Just because of unusual number of tables stacked at end
%%\begin{thebibliography}{99}

%%\bibitem[{\citenamefont{Bird}(1995)}]{bird1995molecular}
%%\bibinfo{author}{\bibfnamefont{G.}~\bibnamefont{Bird}},
%%  \emph{\bibinfo{title}{Molecular gas dynamics and the direct simulation of gas
%%  flows}}, Oxford engineering science series (\bibinfo{publisher}{Clarendon
%%  Press}, \bibinfo{year}{1995}), ISBN \bibinfo{isbn}{9780198561958}.

%%\end{thebibliography}

\clearpage
\renewcommand\thefigure{\arabic{table}}

\begin{table}[!htbp]\tiny
\caption{Vortex center: Vorticity, stream function and location in cavity flow.}\label{table:Table1}
\centering
\begin{tabular}{l|cccccccccc}
\hline
\multirow{2}*{} &	\multicolumn{4}{c}{Primary vortex} & \multicolumn{3}{c}{left secondary vortex} & \multicolumn{3}{c}{right secondary vortex} \\
\cline{2-11}
 & $\omega$ & $\psi$ & $x$ & $y$ & $\psi \times {10^4}$ & $x$ & $y$ &  $\psi \times {10^3}$ & $x$ & $y$  \\
\hline
IDUGKS & 2.0832 & -0.1179 & 0.5346 & 0.5694 & 2.2117 & 0.0817 & 0.0754 & 1.7423 & 0.8581 & 0.1113 \\
Ghia et al~\cite{ghia1982high} & 2.0497 & -0.1179 & 0.5313 & 0.5625 & 2.3113 & 0.0859 & 0.0781 & 1.7510 & 0.8594 & 0.1094 \\
Error (\%) & 1.60 & 0.00 & 0.33 & 0.69 & 4.30 & 0.42 & 0.27 & 0.50 & 0.13 & 0.19 \\
Hou et al~\cite{hou1995simulation} & 2.0760 & -0.1178 & 0.5333 & 0.5647 & 2.2200 & 0.0902 & 0.0784 & 1.6900 & 0.8667 & 0.1137 \\
Zhuo et al~\cite{zhuo2012filter} & 2.0570 & -0.1179 & 0.5311 & 0.5662 & 2.2667 & 0.0828 & 0.0770 & 1.7066 & 0.8645 & 0.1120 \\

\hline
\end{tabular}\\
\end{table}

\begin{table}[!htbp]\tiny
\caption{Comparison for computational cost between the IDUGKS and explicit DUGKS.}\label{table:Table2}
\centering
\begin{tabular}{l|rrrr|r}
\hline
\multirow{2}*{} & \multicolumn{2}{c}{Explicit method} &  \multicolumn{2}{c|}{Implicit method} & \multirow{2}*{Speedup} \\
\cline{2-5}
 &  iteration steps & time(s) & iteration steps & time(s) \\
\hline
Lid-driven Cavity flow under uniform mesh(Re=1000) & 255000 & 45390.4 & 2040 & 504.6 & 89.95\\
Lid-driven Cavity flow under hybrid mesh(Re=1000) & 240000 & 5716.6 & 1750 & 75.3 & 75.9\\
Lid-driven Cavity flow(Kn=0.075) & 11900 & 4188.8 & 230 & 152.4 & 27.49\\
Lid-driven Cavity flow (Kn=1.0) & 9600 & 3479.3 & 190 & 129.8 & 26.80\\
Lid-driven Cavity flow (Kn=10) & 34200 & 11938.9 & 640 & 418.6 & 28.52\\
Hypersonic cylinder case under structured grid(Kn=1.0) & 18100 & 18955.6 & 310 & 616.2 & 30.76\\
Hypersonic cylinder case under hybrid mesh(Kn=1.0) & 16400 & 15611.2 & 300 & 542.4 & 28.7\\
\hline
\end{tabular}\\
\end{table}

\clearpage
\renewcommand\thefigure{\arabic{figure}}

\begin{figure}
  \centering
  \includegraphics[width=0.5\textwidth]{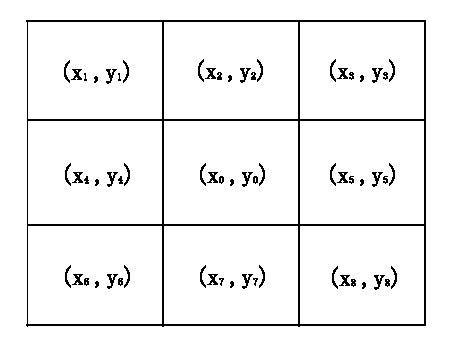}
  \caption{Cells for reconstruction.}
  \label{fig:Fig01}
\end{figure}

\begin{figure}
  \centering
  \includegraphics[width=0.5\textwidth]{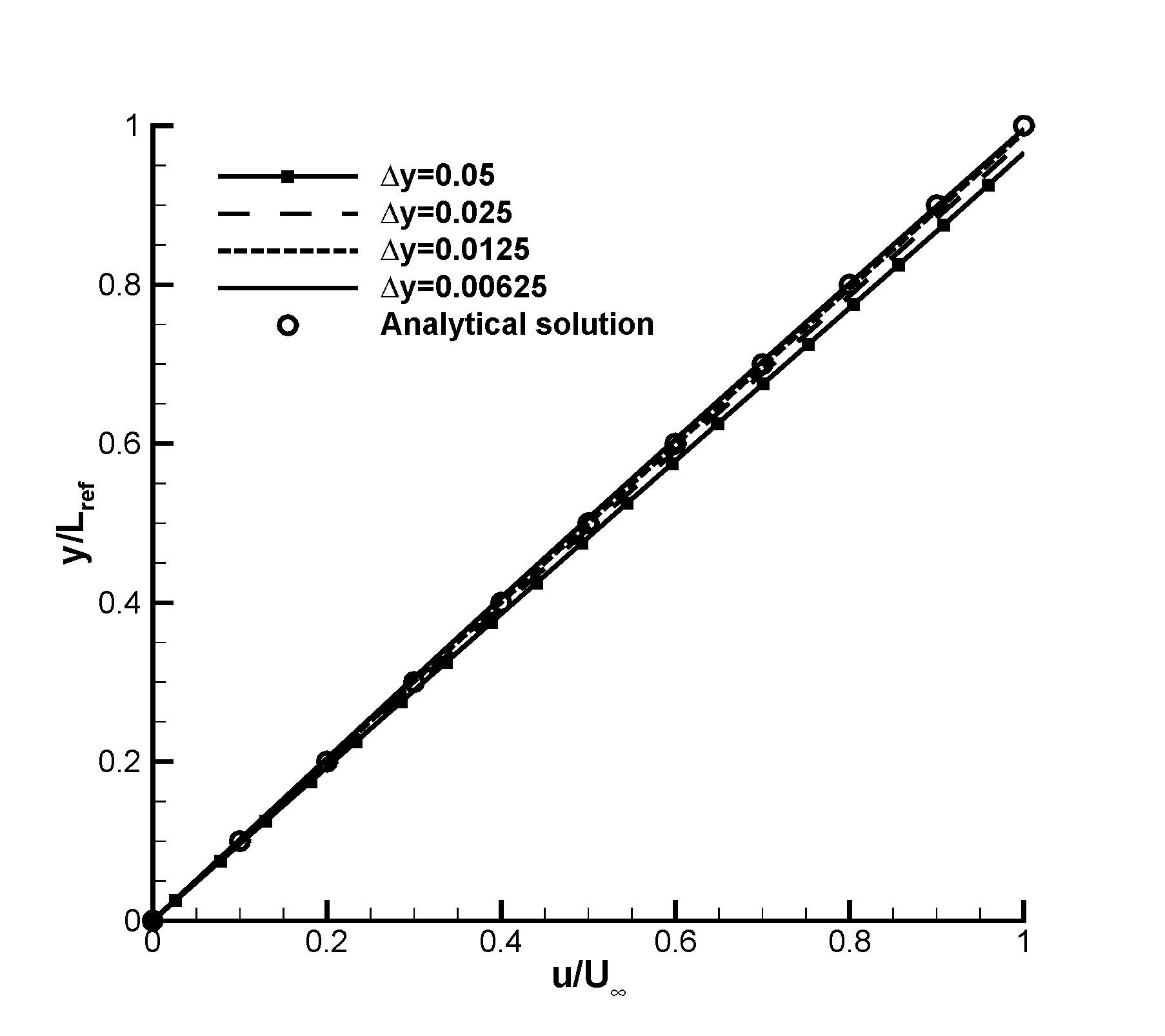}
  \caption{The comparison result for U velocity profiles of Couette flow on different meshes.}
  \label{fig:Fig02}
\end{figure}

\begin{figure}
  \centering
  \includegraphics[width=0.5\textwidth]{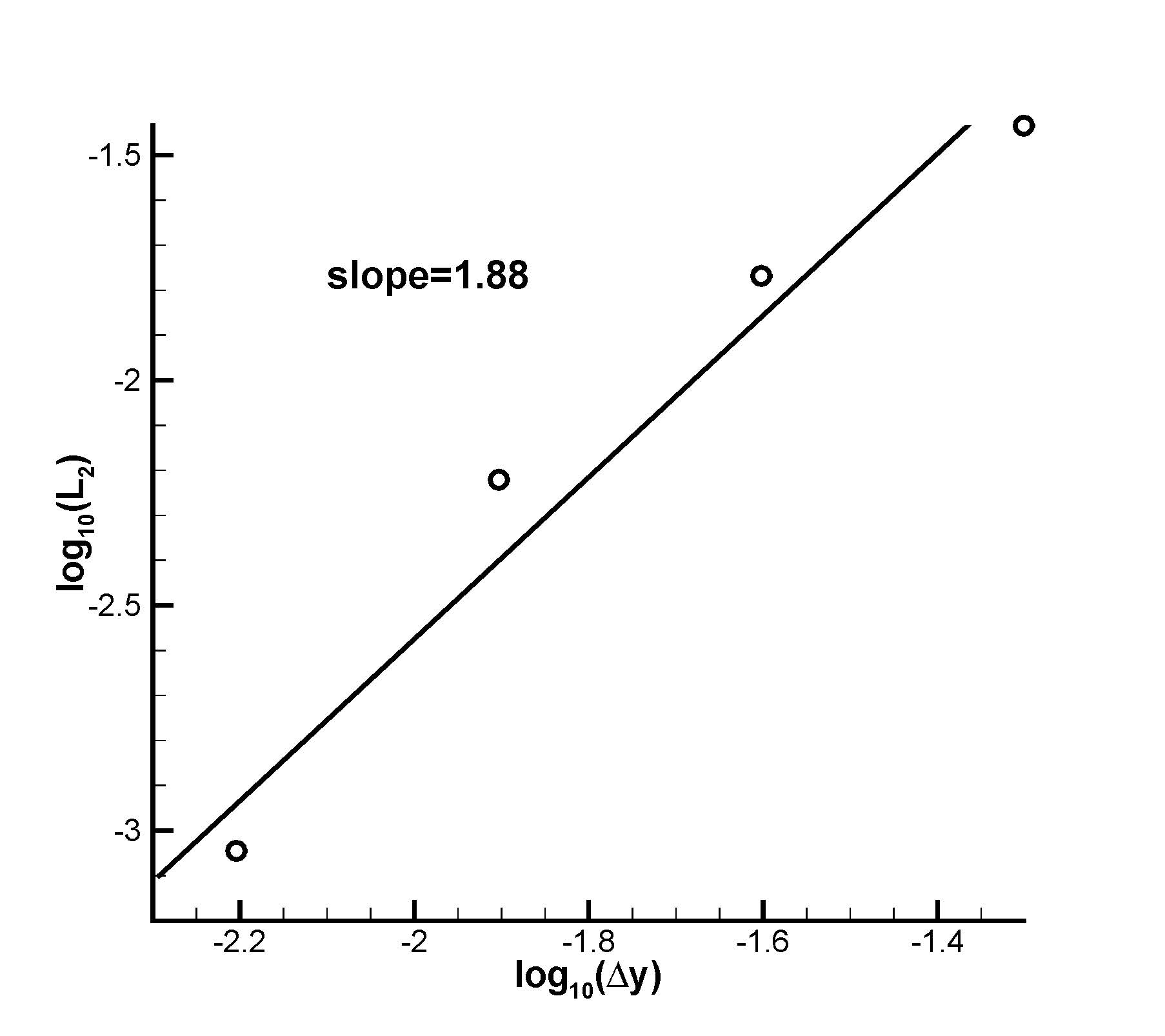}
  \caption{The L2 norm of absolute errors on different meshes for the Couette flow.}
  \label{fig:Fig03}
\end{figure}

\begin{figure}
  \centering
  \includegraphics[width=0.5\textwidth]{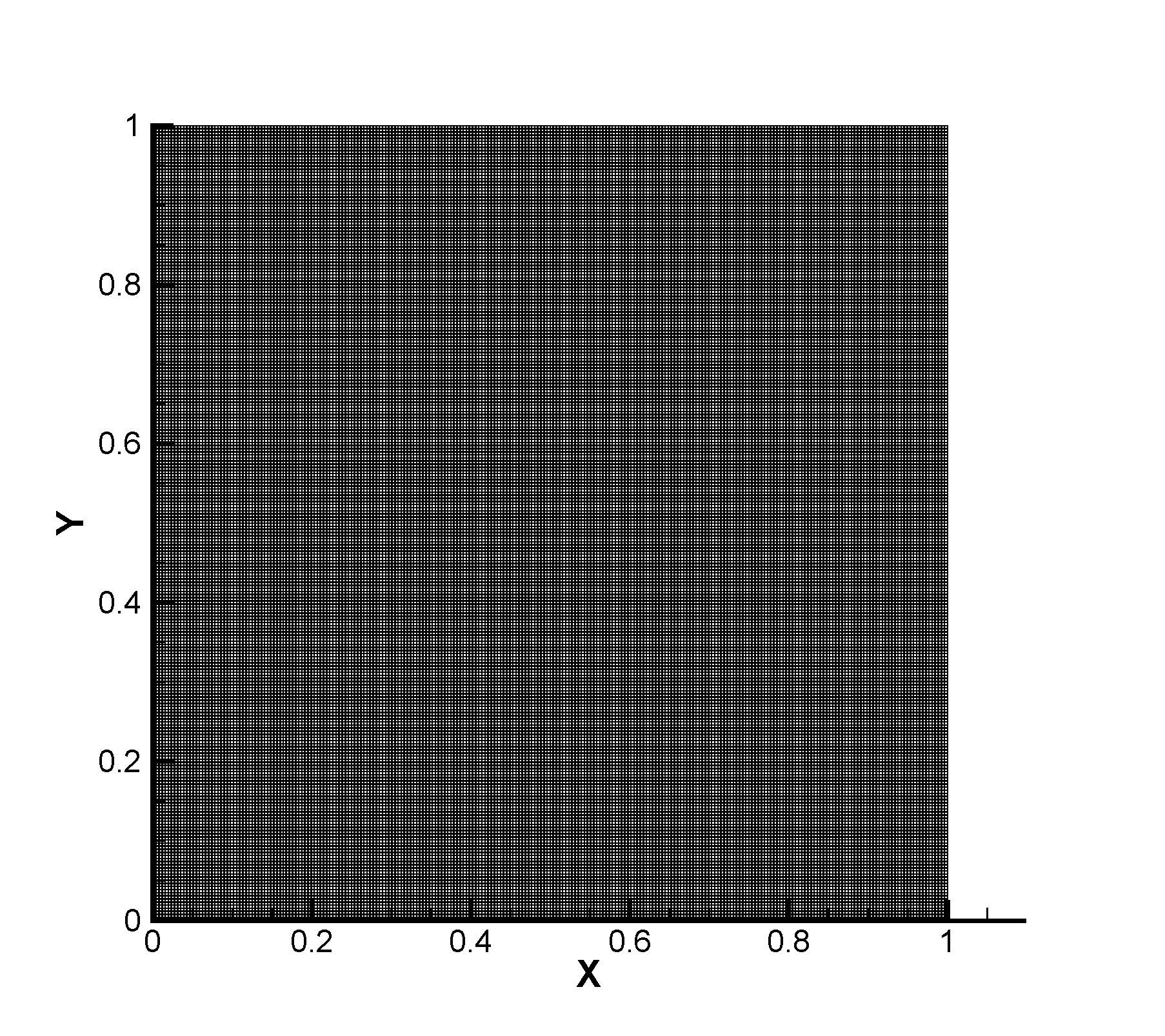}
  \caption{Uniform mesh for incompressible lid-driven cavity flow.}
  \label{fig:Fig04}
\end{figure}

\begin{figure}
  \centering
  \includegraphics[width=0.5\textwidth]{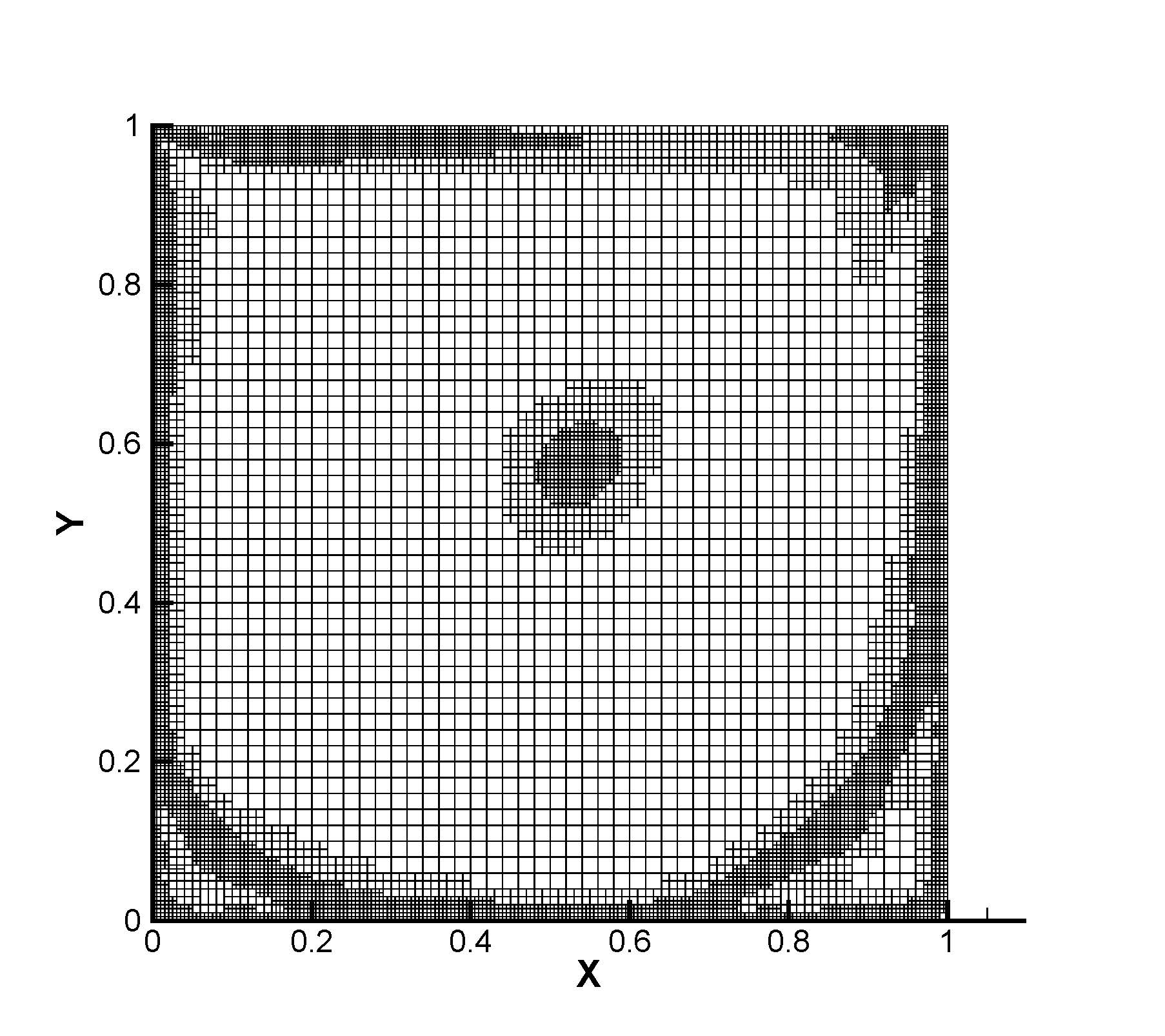}
  \caption{Hybrid mesh for incompressible lid-driven cavity flow.}
  \label{fig:Fig05}
\end{figure}

\begin{figure}
  \centering
  \includegraphics[width=0.5\textwidth]{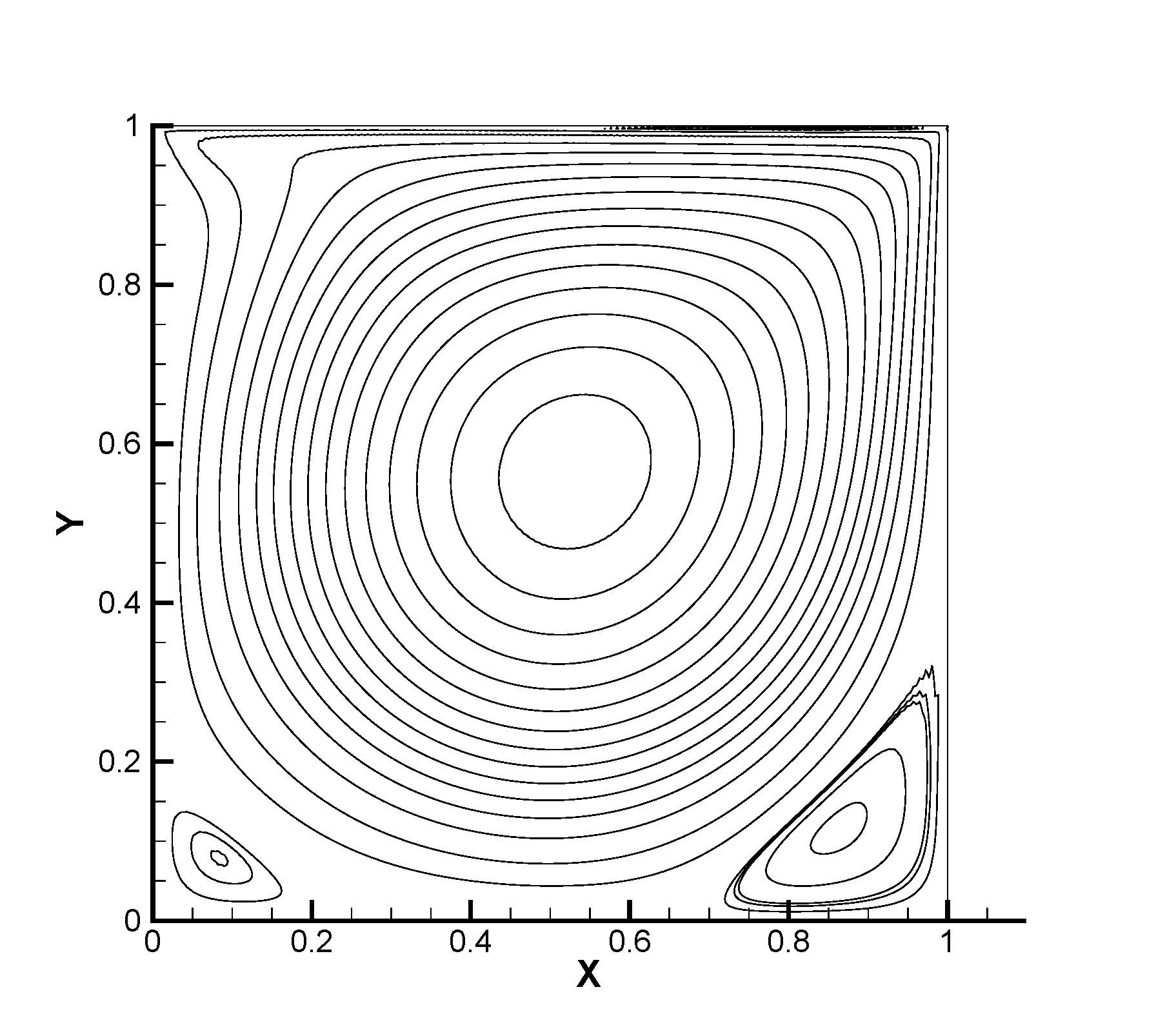}
  \caption{The streamlines of lid-driven cavity flow ($Re = 1000$).}
  \label{fig:Fig06}
\end{figure}

\begin{figure}
  \centering
  \includegraphics[width=0.5\textwidth]{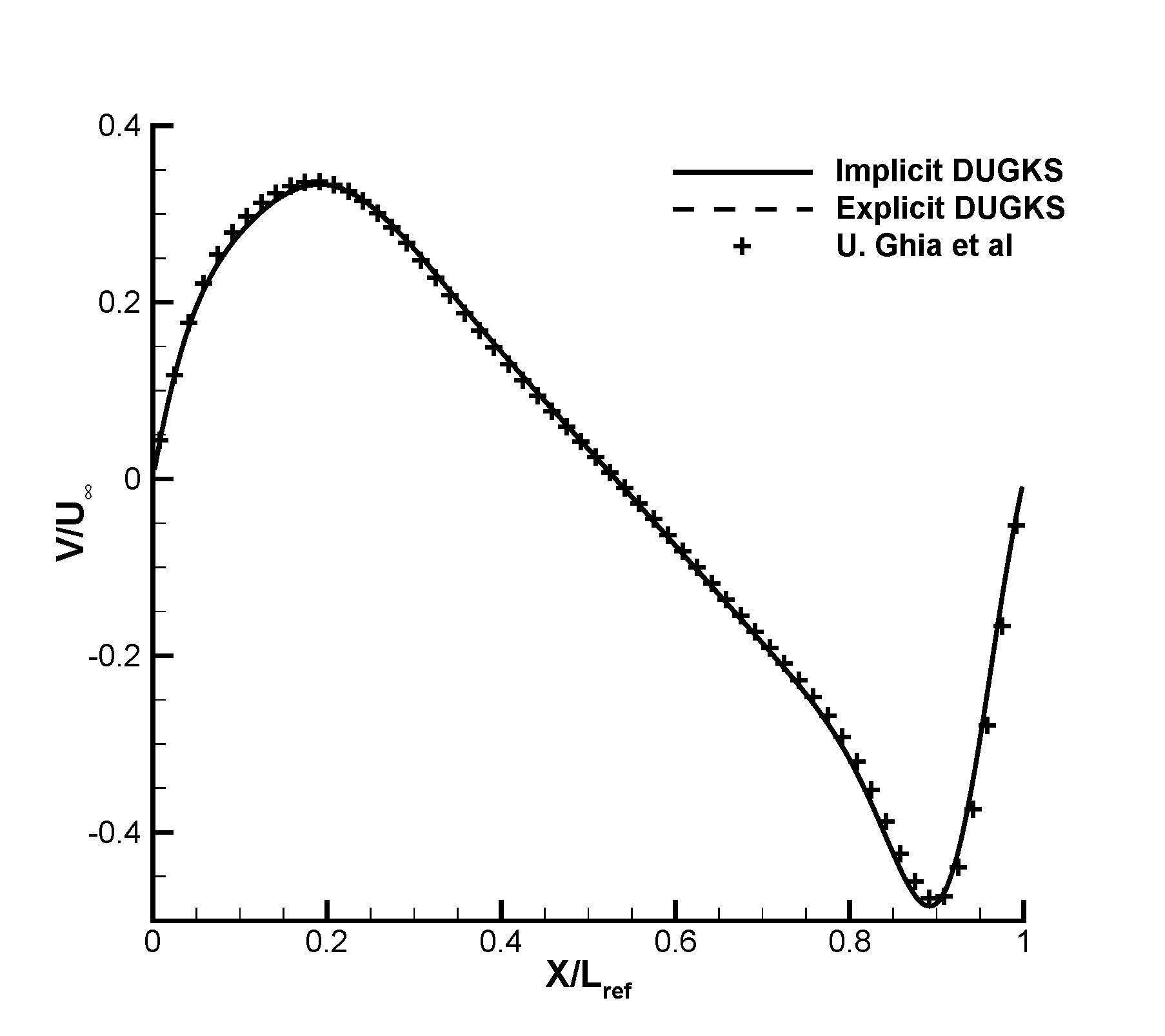}
  \caption{Comparison result for V velocity profile of lid-driven cavity flow ($Re = 1000$).}
  \label{fig:Fig07}
\end{figure}

\begin{figure}
  \centering
  \includegraphics[width=0.5\textwidth]{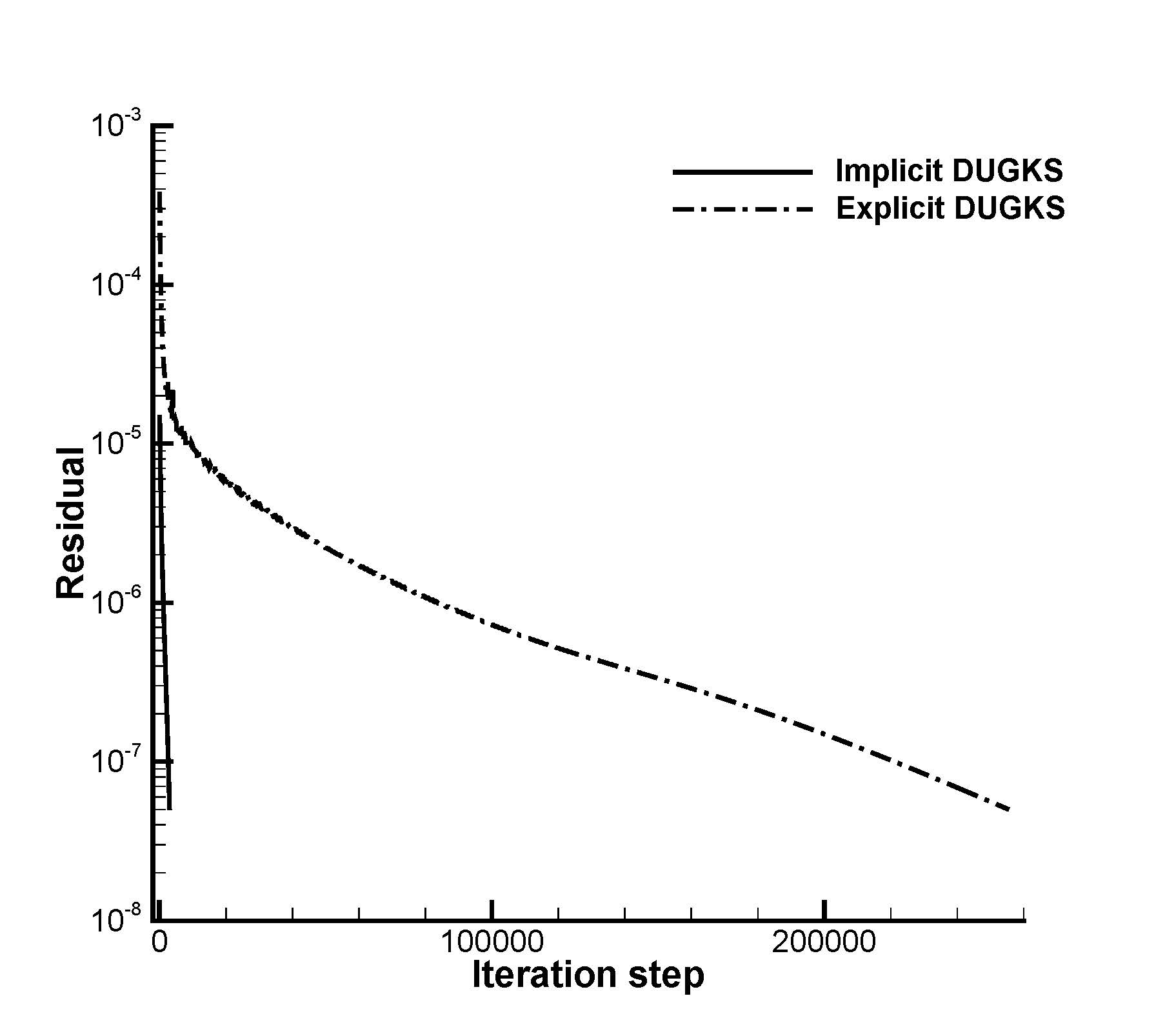}
  \caption{The residual curves of lid-driven cavity flow ($Re = 1000$).}
  \label{fig:Fig08}
\end{figure}

\begin{figure}
  \centering
 \subfigure[U profile at line $X = 0.5$]{\label{fig:Fig09a}\includegraphics[width=0.45\textwidth]{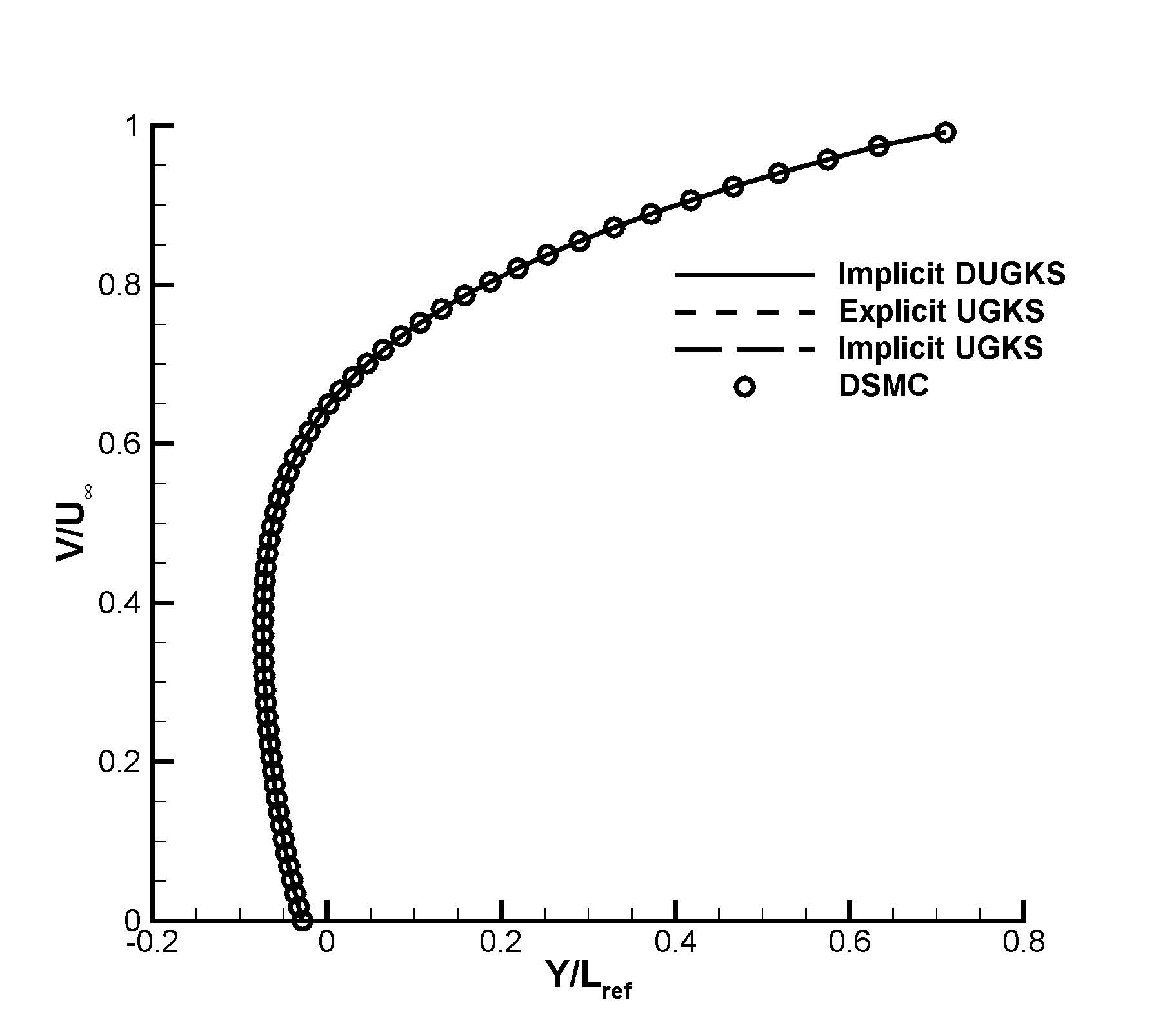}}
 \subfigure[V profile at line $Y = 0.5$]{\label{fig:Fig09b}\includegraphics[width=0.45\textwidth]{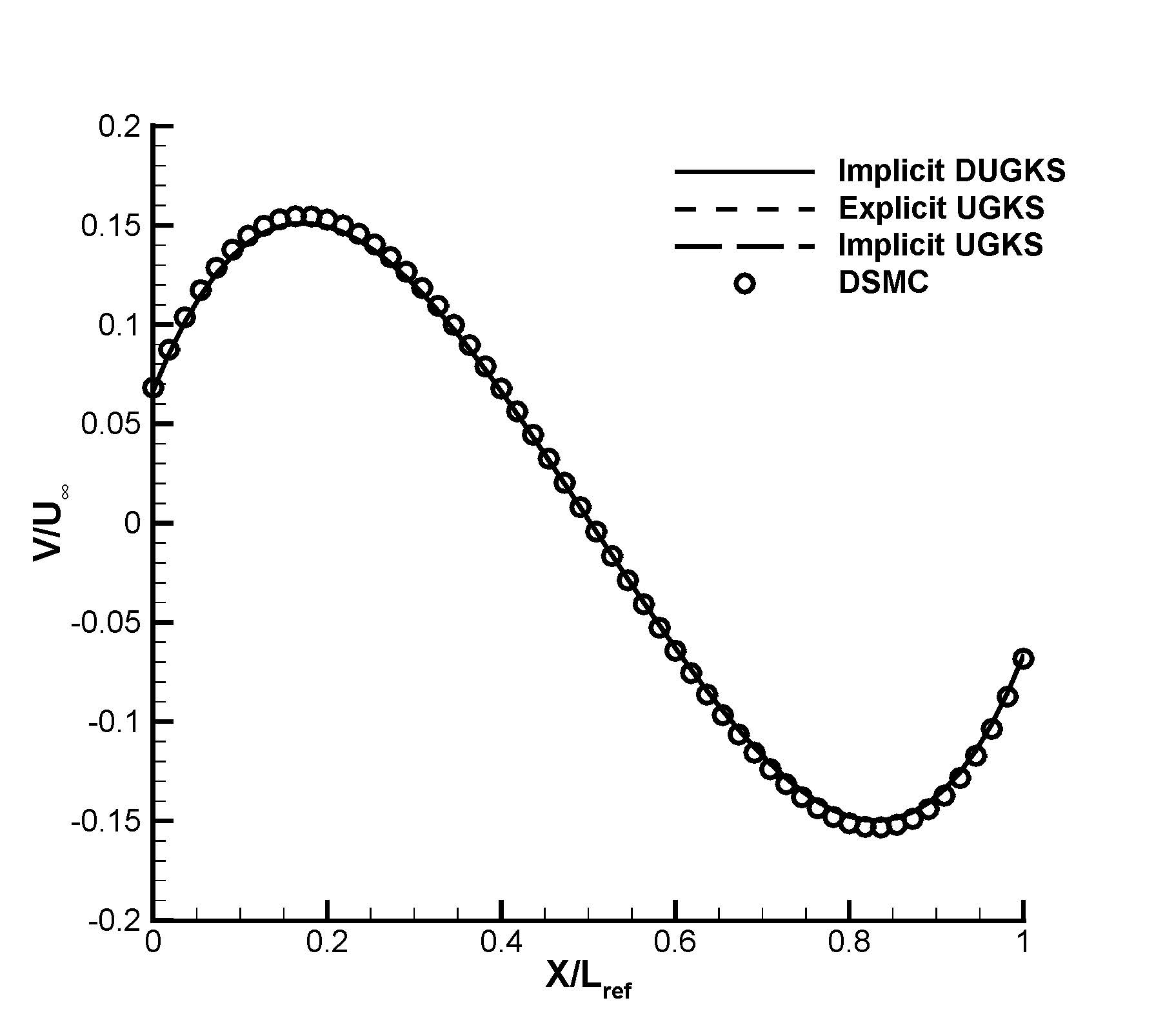}}
  \caption{Comparison results for velocity profiles of lid-driven cavity flow ($Kn = 0.075$).}
  \label{fig:Fig09}
\end{figure}

\begin{figure}
  \centering
  \includegraphics[width=0.5\textwidth]{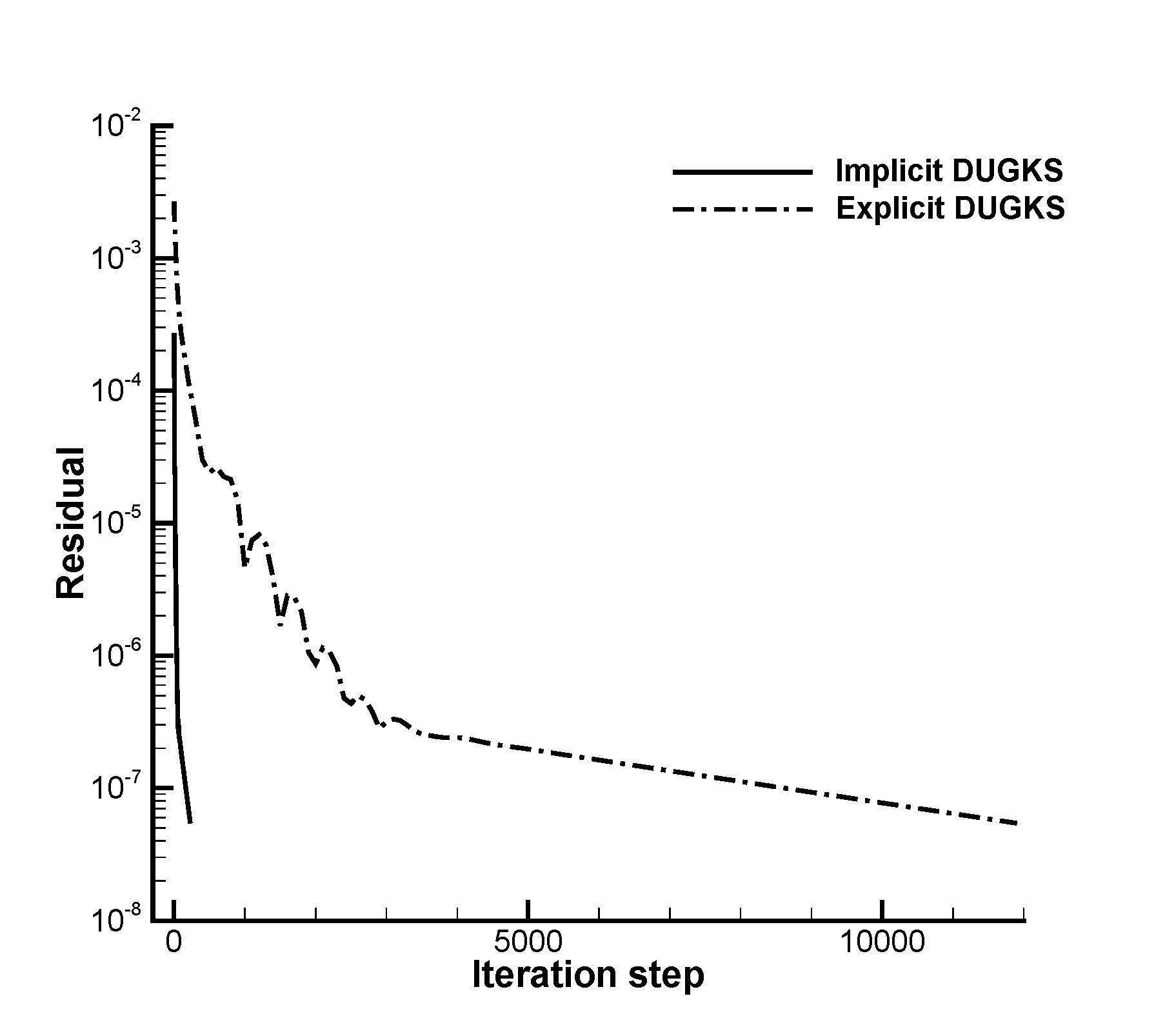}
  \caption{The residual curves of lid-driven cavity flow ($Kn = 0.075$).}
  \label{fig:Fig10}
\end{figure}

\begin{figure}
  \centering
 \subfigure[U profile at line $X = 0.5$]{\label{fig:Fig11a}\includegraphics[width=0.45\textwidth]{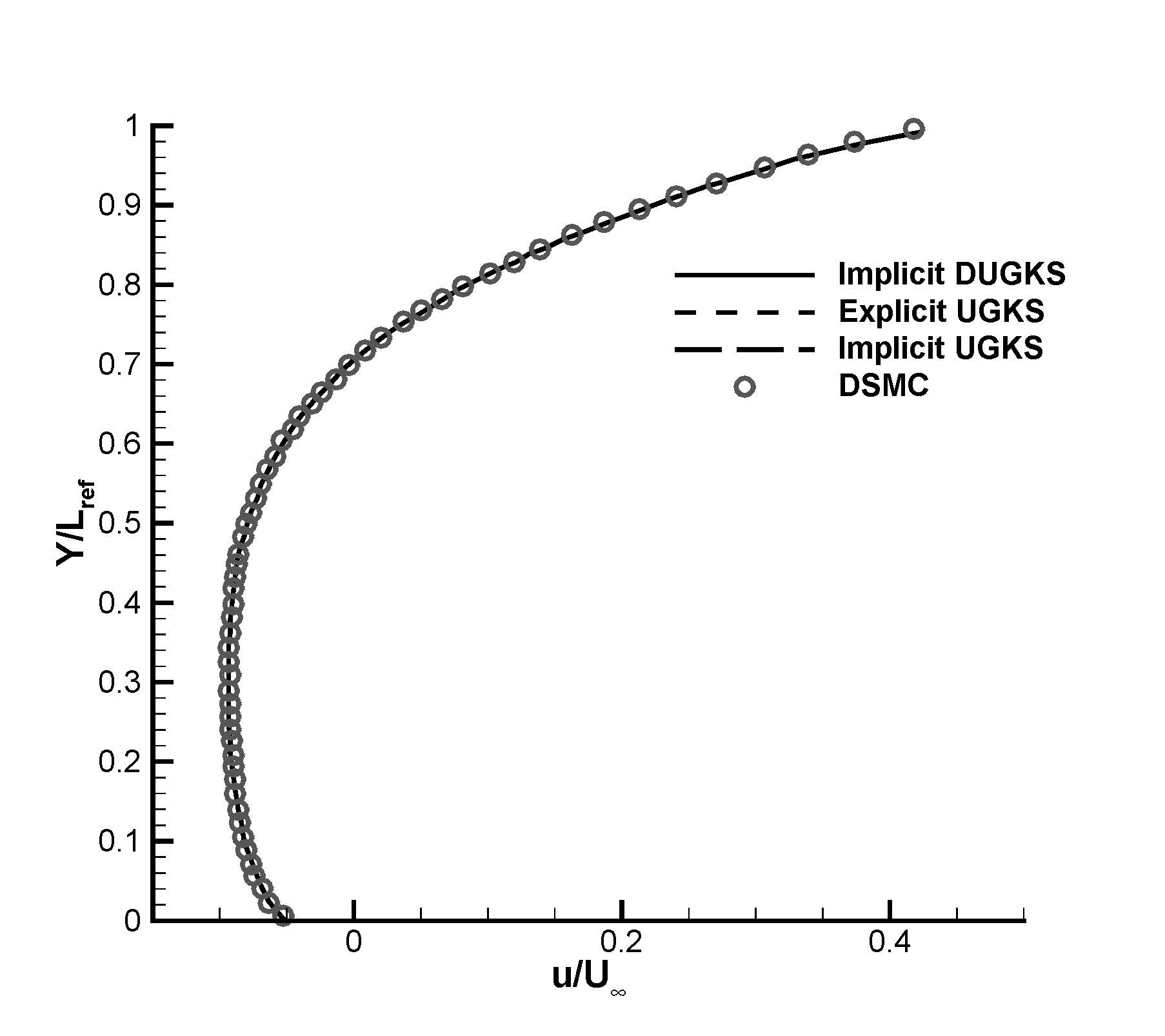}}
 \subfigure[V profile at line $Y = 0.5$]{\label{fig:Fig11b}\includegraphics[width=0.45\textwidth]{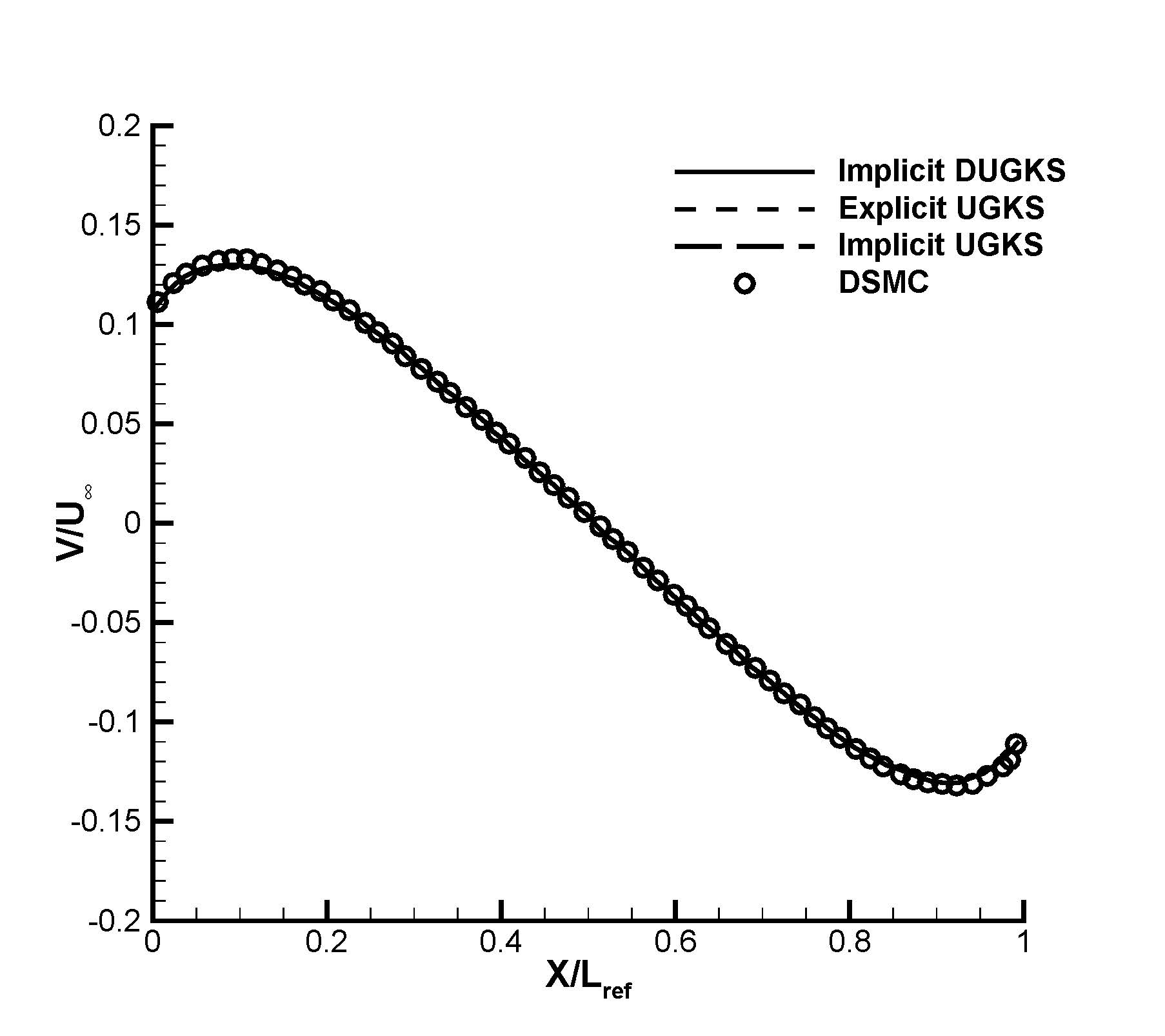}}
  \caption{Comparison results for velocity profiles of lid-driven cavity flow ($Kn = 1.0$).}
  \label{fig:Fig11}
\end{figure}

\begin{figure}
  \centering
  \includegraphics[width=0.5\textwidth]{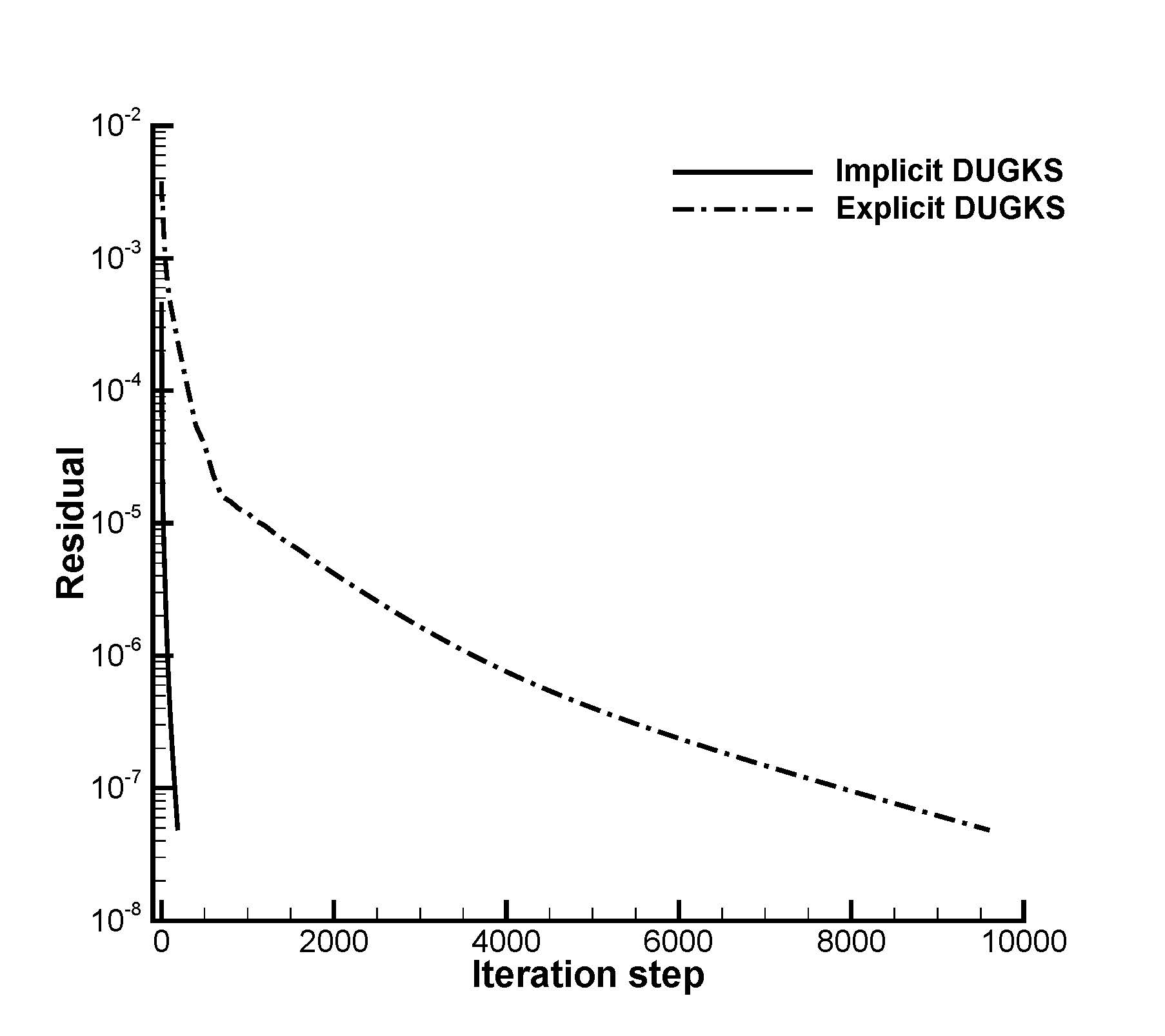}
  \caption{The residual curves of lid-driven cavity flow ($Kn = 1.0$).}
  \label{fig:Fig12}
\end{figure}

\begin{figure}
  \centering
 \subfigure[U profile at line $X = 0.5$]{\label{fig:Fig13a}\includegraphics[width=0.45\textwidth]{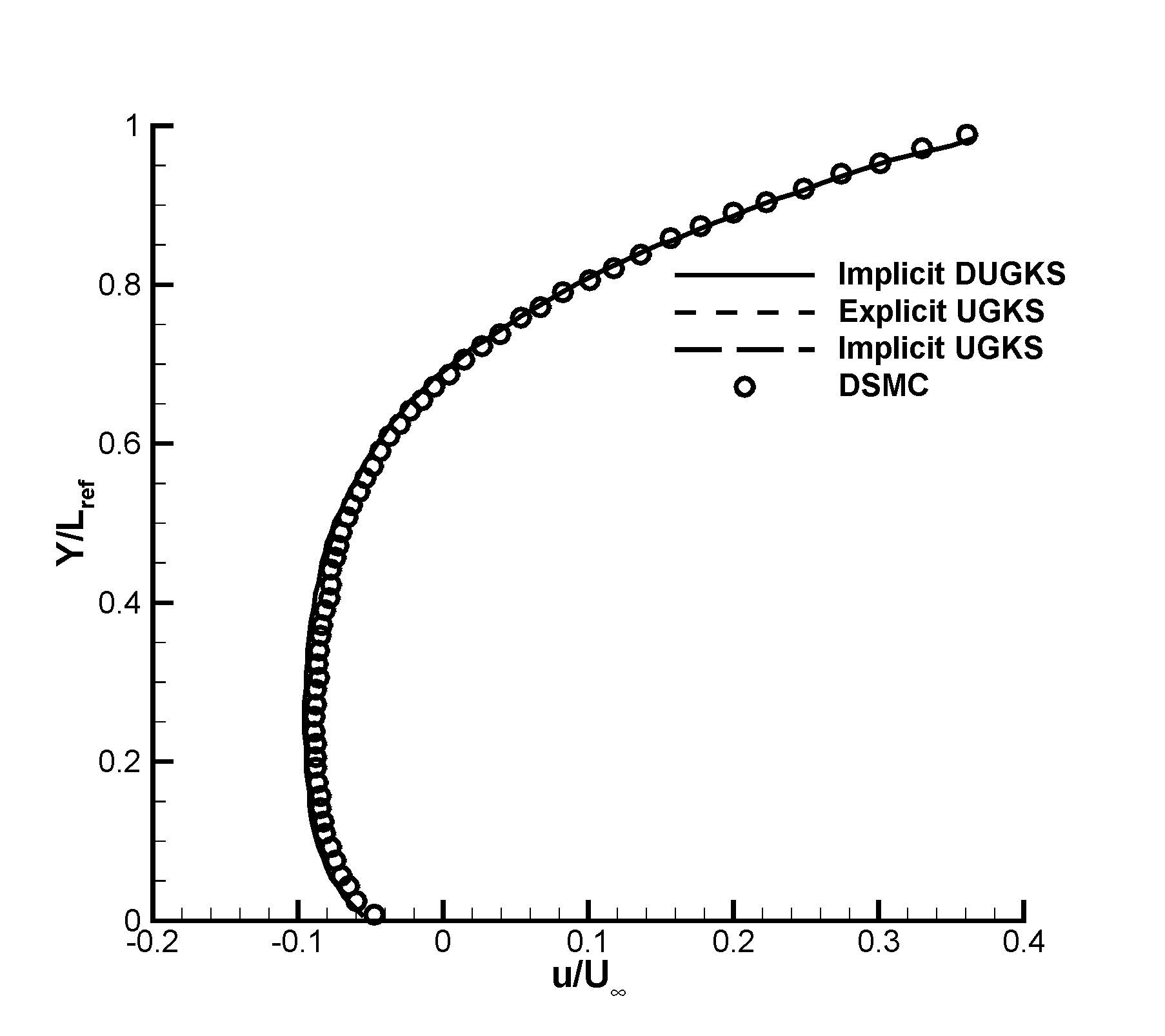}}
 \subfigure[V profile at line $Y = 0.5$]{\label{fig:Fig13b}\includegraphics[width=0.45\textwidth]{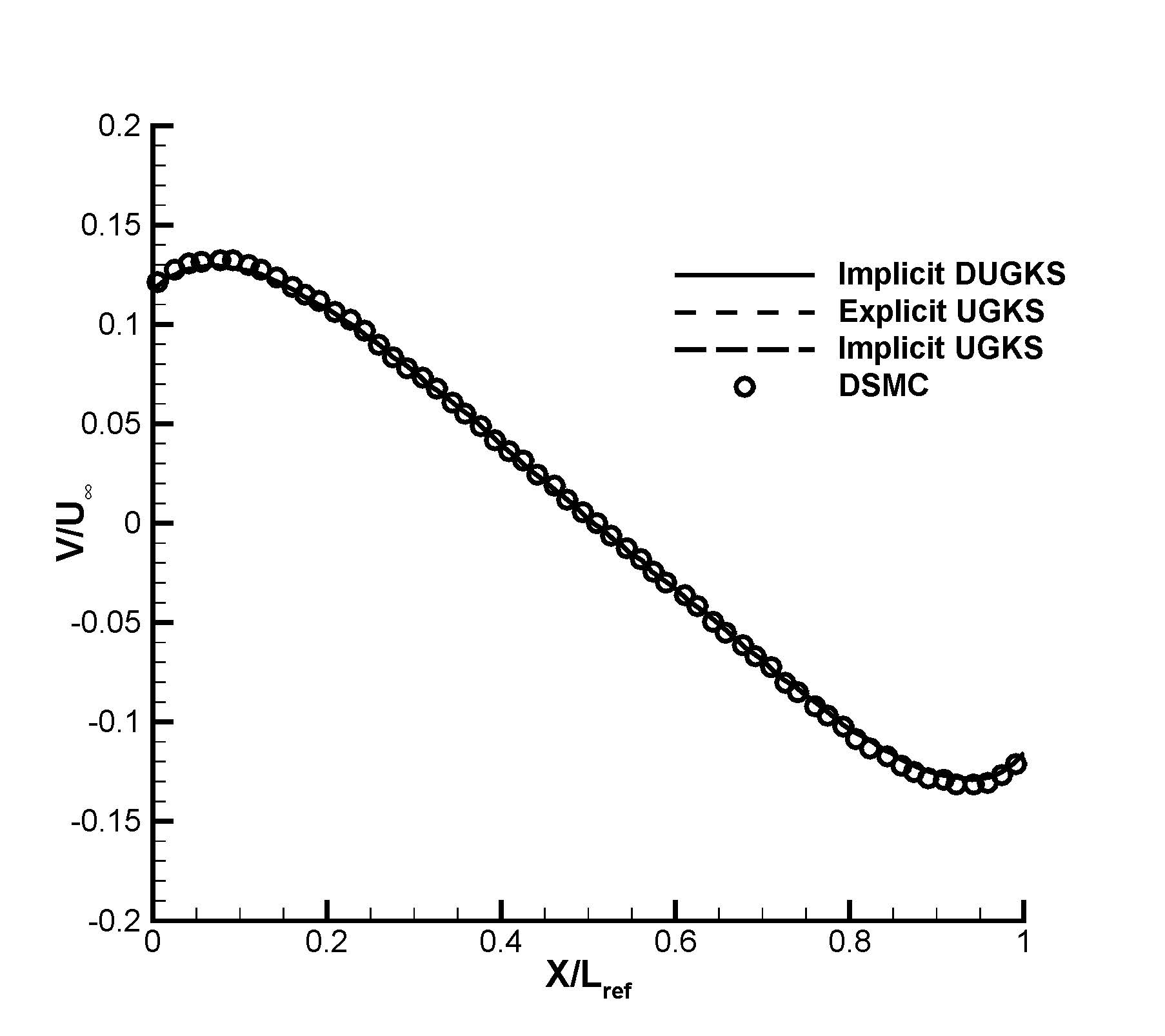}}
  \caption{Comparison results for velocity profiles of lid-driven cavity flow ($Kn = 10.0$).}
  \label{fig:Fig13}
\end{figure}

\begin{figure}
  \centering
  \includegraphics[width=0.5\textwidth]{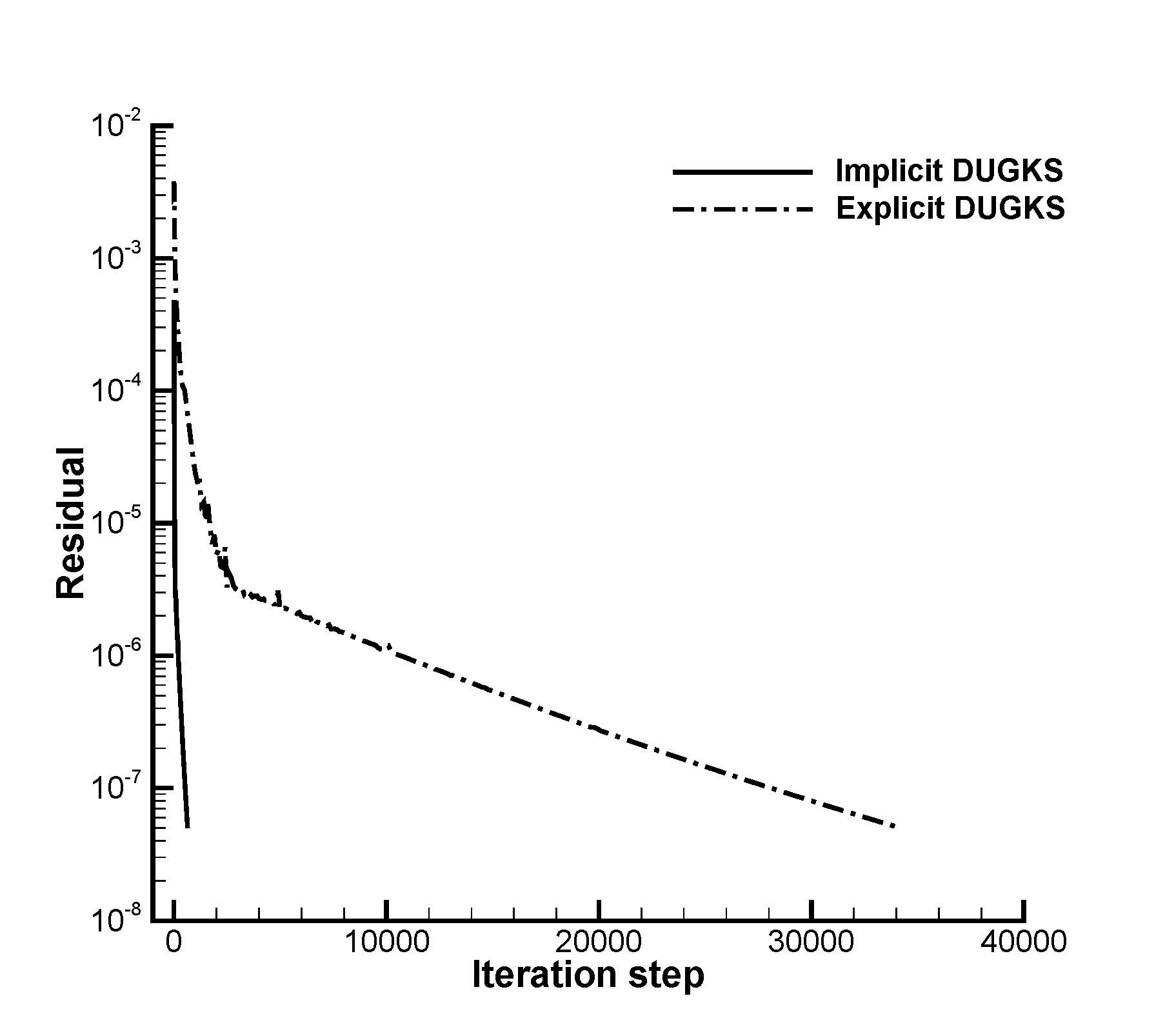}
  \caption{The residual curves of lid-driven cavity flow ($Kn = 10.0$).}
  \label{fig:Fig14}
\end{figure}

\begin{figure}
  \centering
 \subfigure[Density contour]{\label{fig:Fig15a}\includegraphics[width=0.45\textwidth]{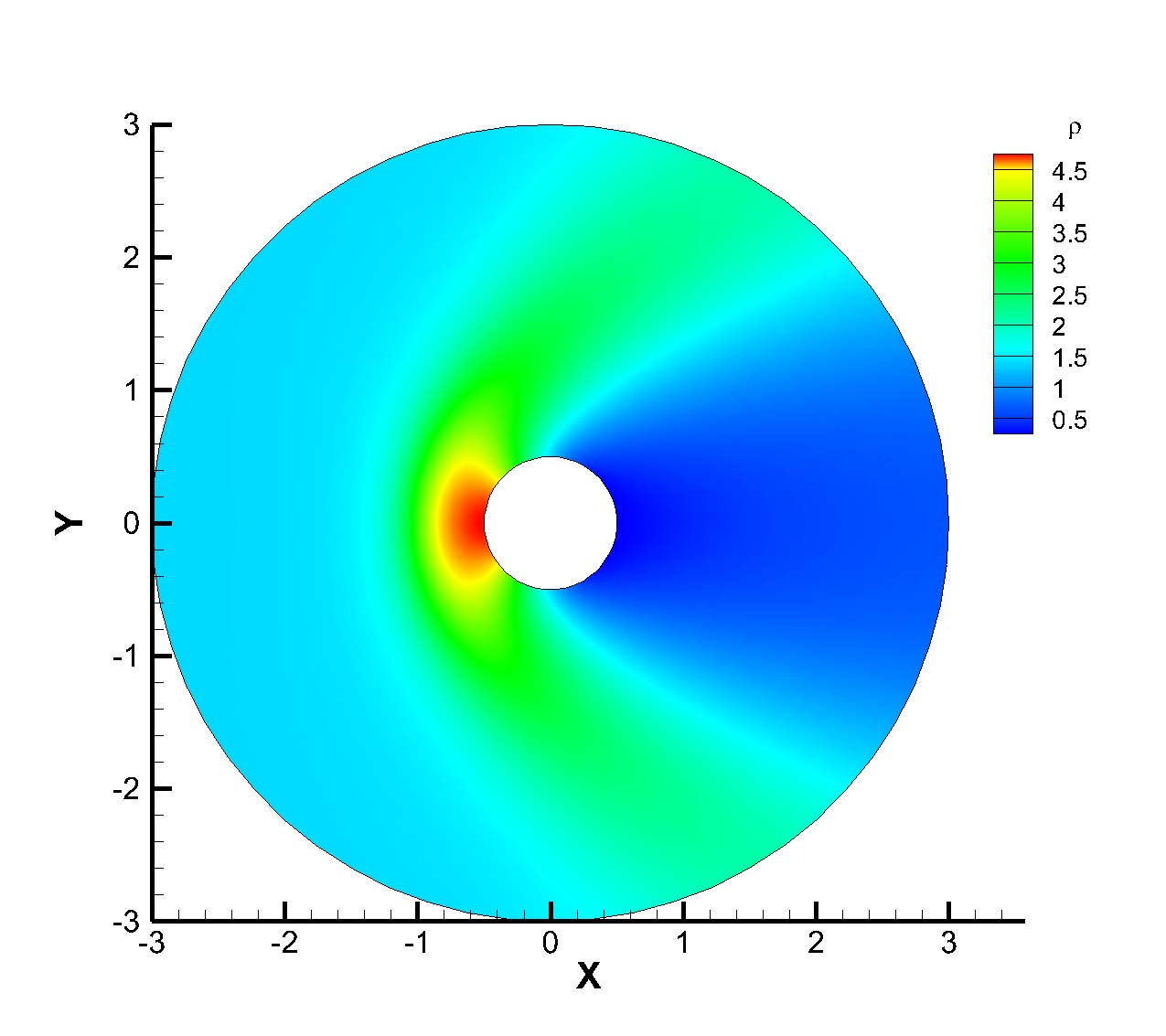}}
 \subfigure[Pressure contour]{\label{fig:Fig15b}\includegraphics[width=0.45\textwidth]{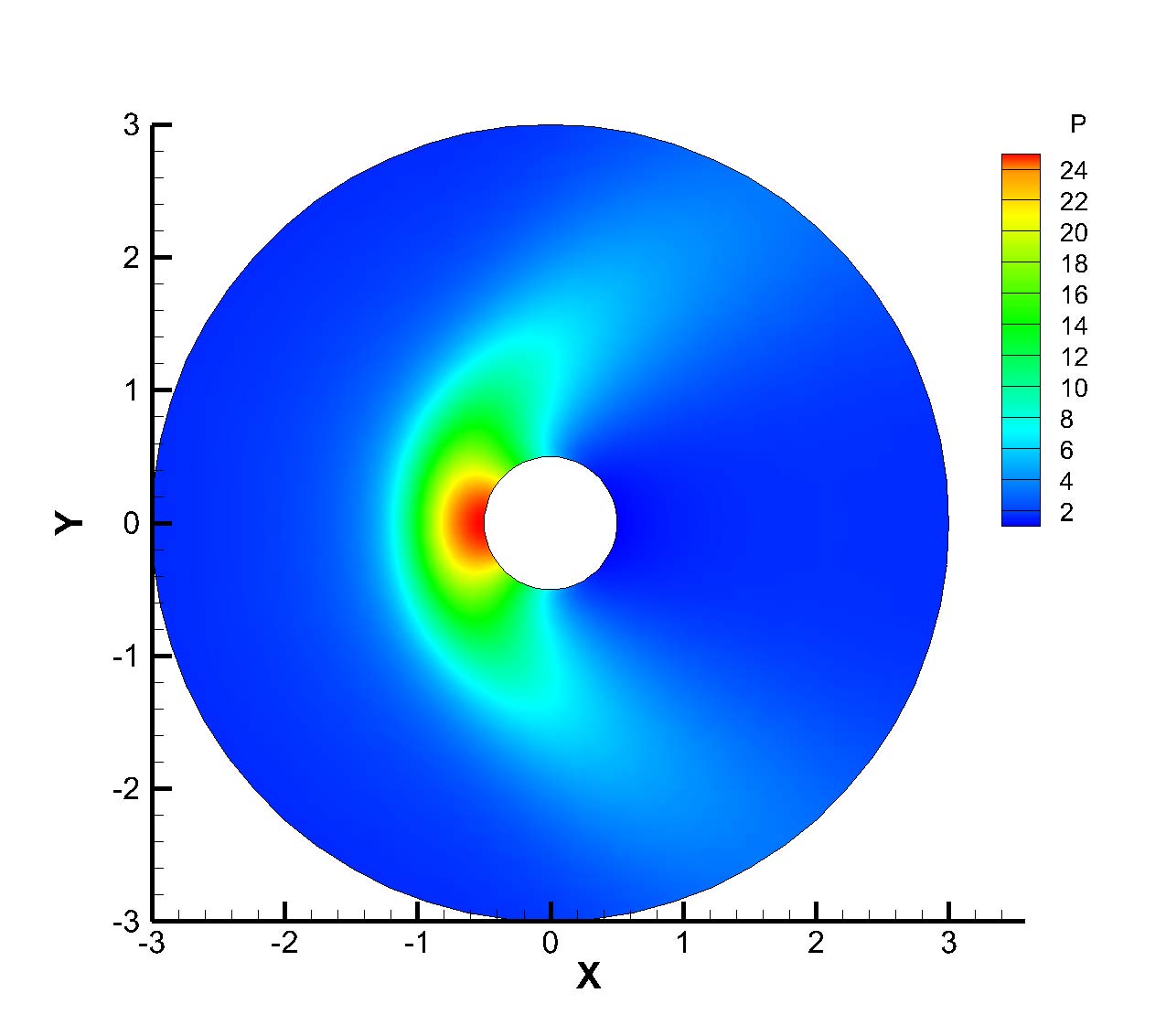}}
  \caption{Hypersonic flow around a circular cylinder.}
  \label{fig:Fig15}
\end{figure}

\begin{figure}
  \centering
  \includegraphics[width=0.5\textwidth]{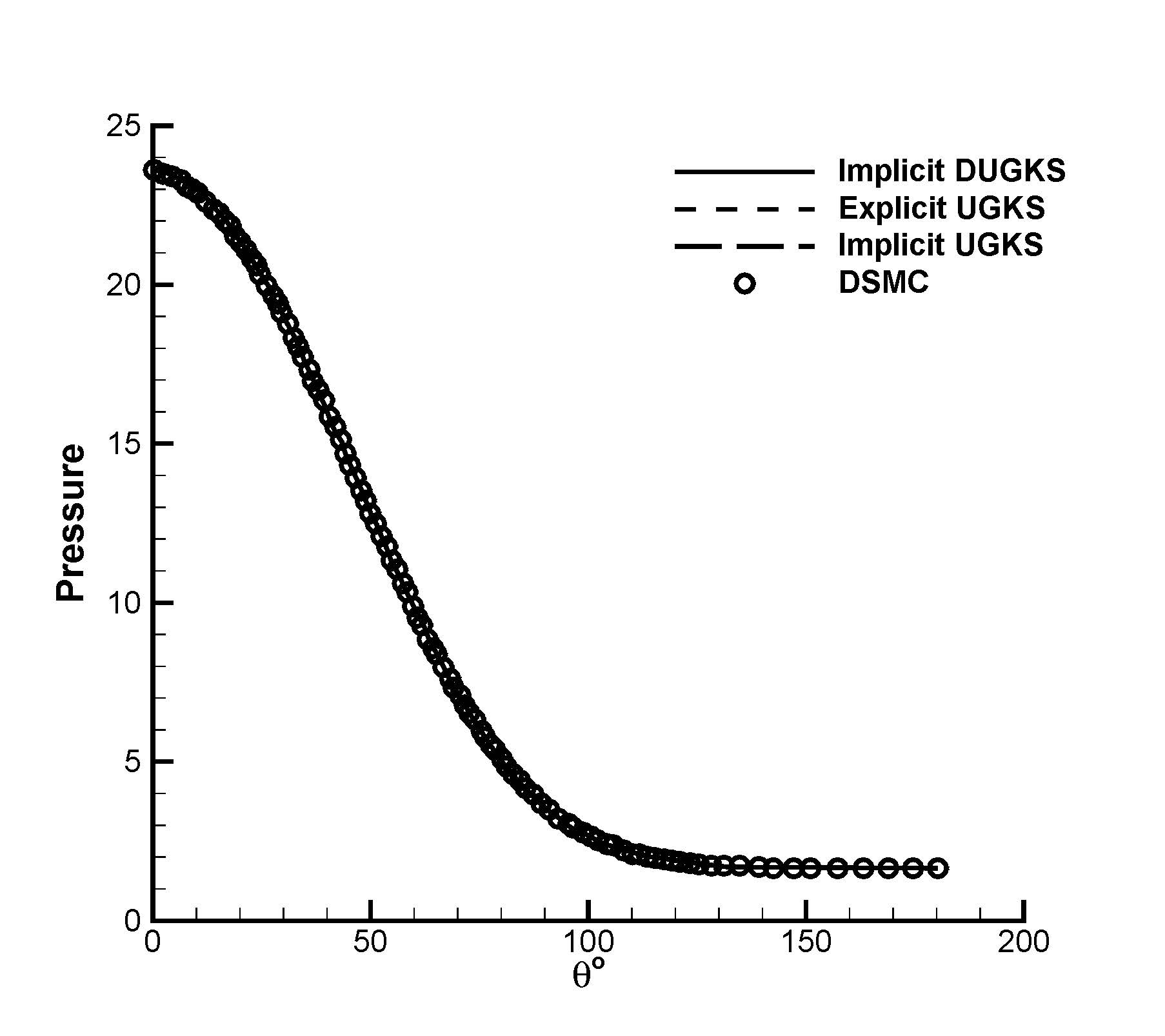}
  \caption{The pressure distribution on surface of circular cylinder.}
  \label{fig:Fig16}
\end{figure}

\begin{figure}
  \centering
  \includegraphics[width=0.5\textwidth]{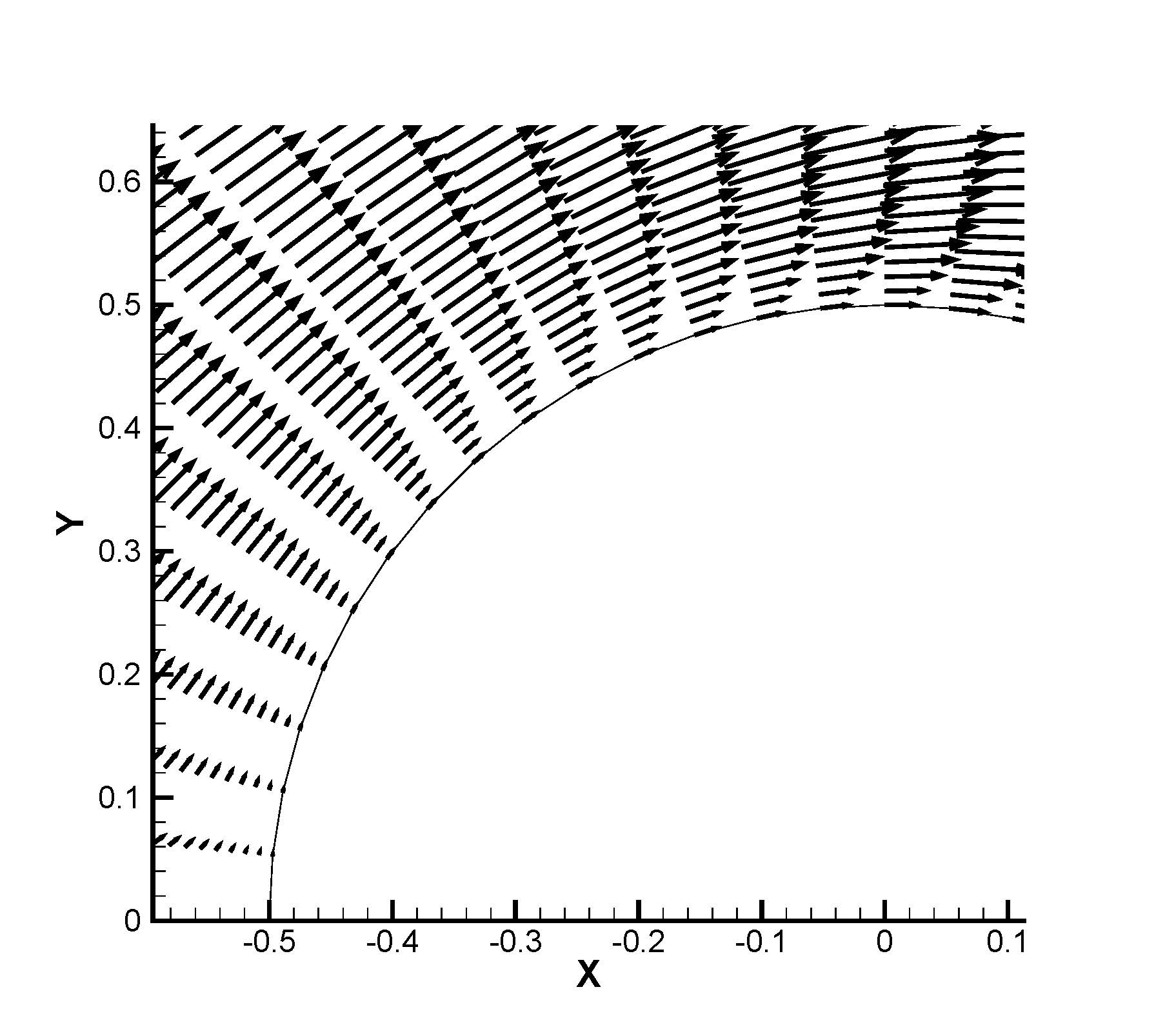}
  \caption{The velocity vectors near circular cylinder.}
  \label{fig:Fig17}
\end{figure}

\begin{figure}
  \centering
  \includegraphics[width=0.5\textwidth]{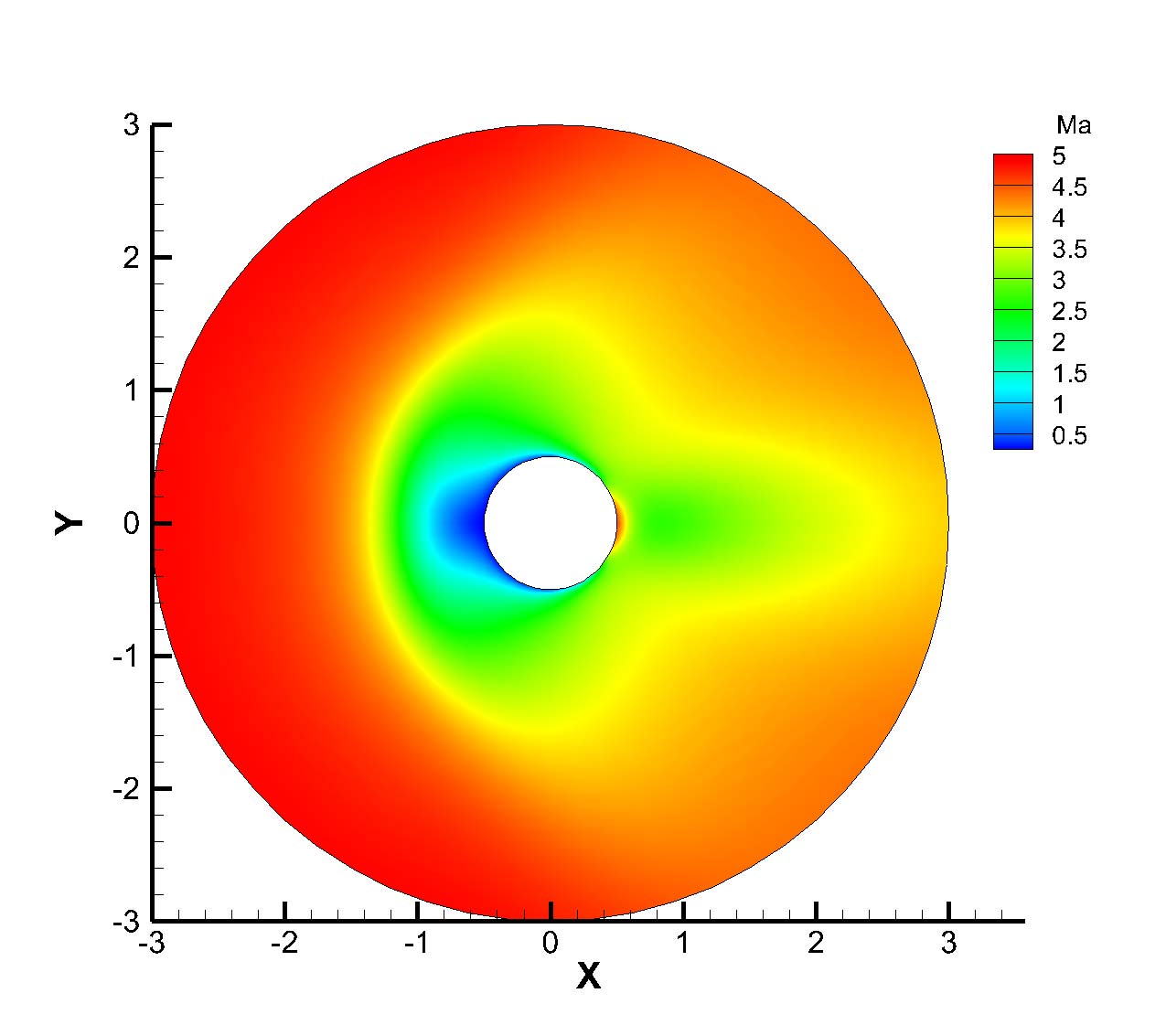}
  \caption{The Mach number contour of hypersonic circular cylinder flow.}
  \label{fig:Fig18}
\end{figure}

\clearpage

\begin{figure}
  \centering
  \includegraphics[width=0.5\textwidth]{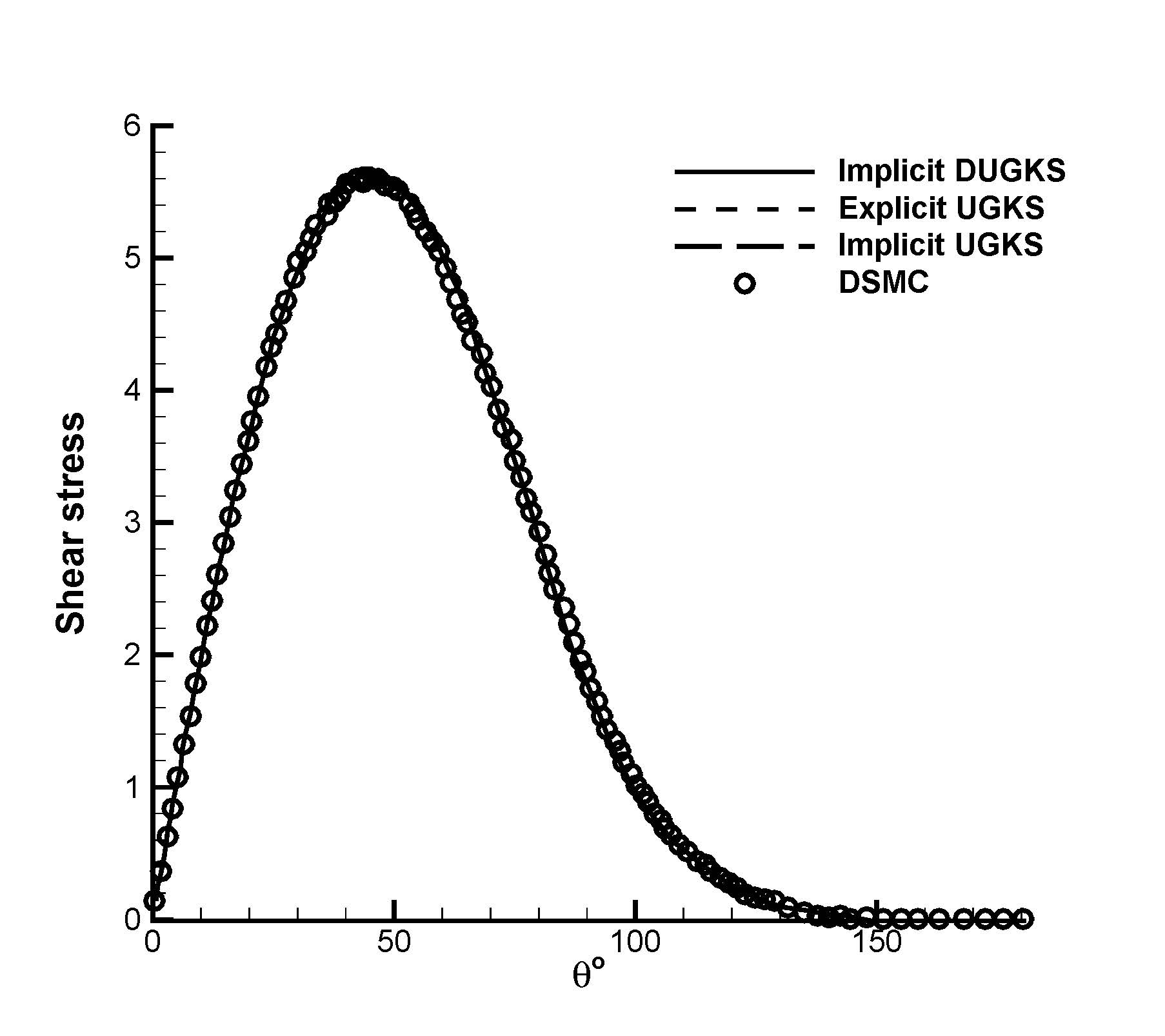}
  \caption{The shear stress distribution on surface of circular cylinder.}
  \label{fig:Fig19}
\end{figure}

\begin{figure}
  \centering
  \includegraphics[width=0.5\textwidth]{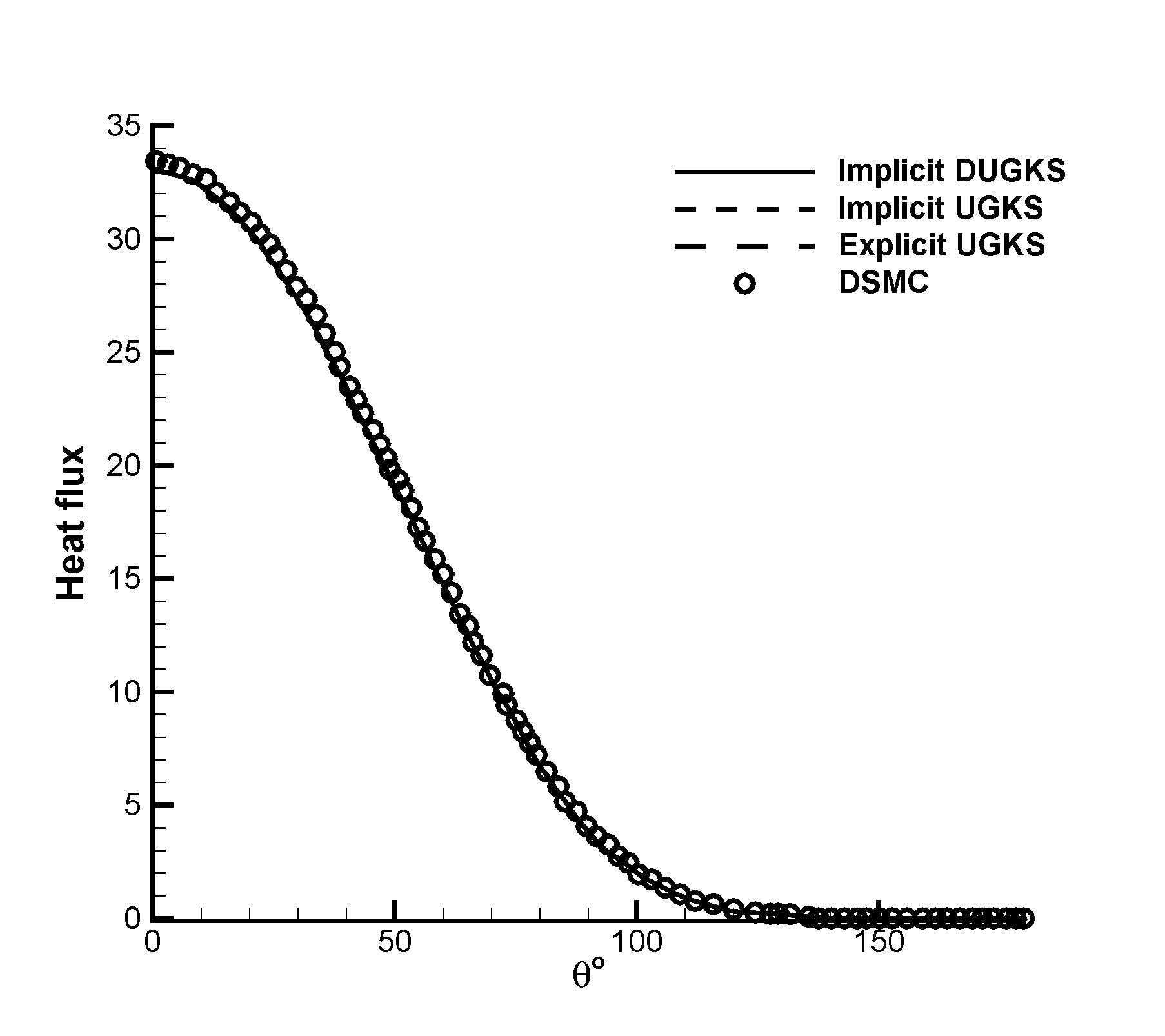}
  \caption{The heat flux distribution on surface of circular cylinder.}
  \label{fig:Fig20}
\end{figure}

\begin{figure}
  \centering
  \includegraphics[width=0.5\textwidth]{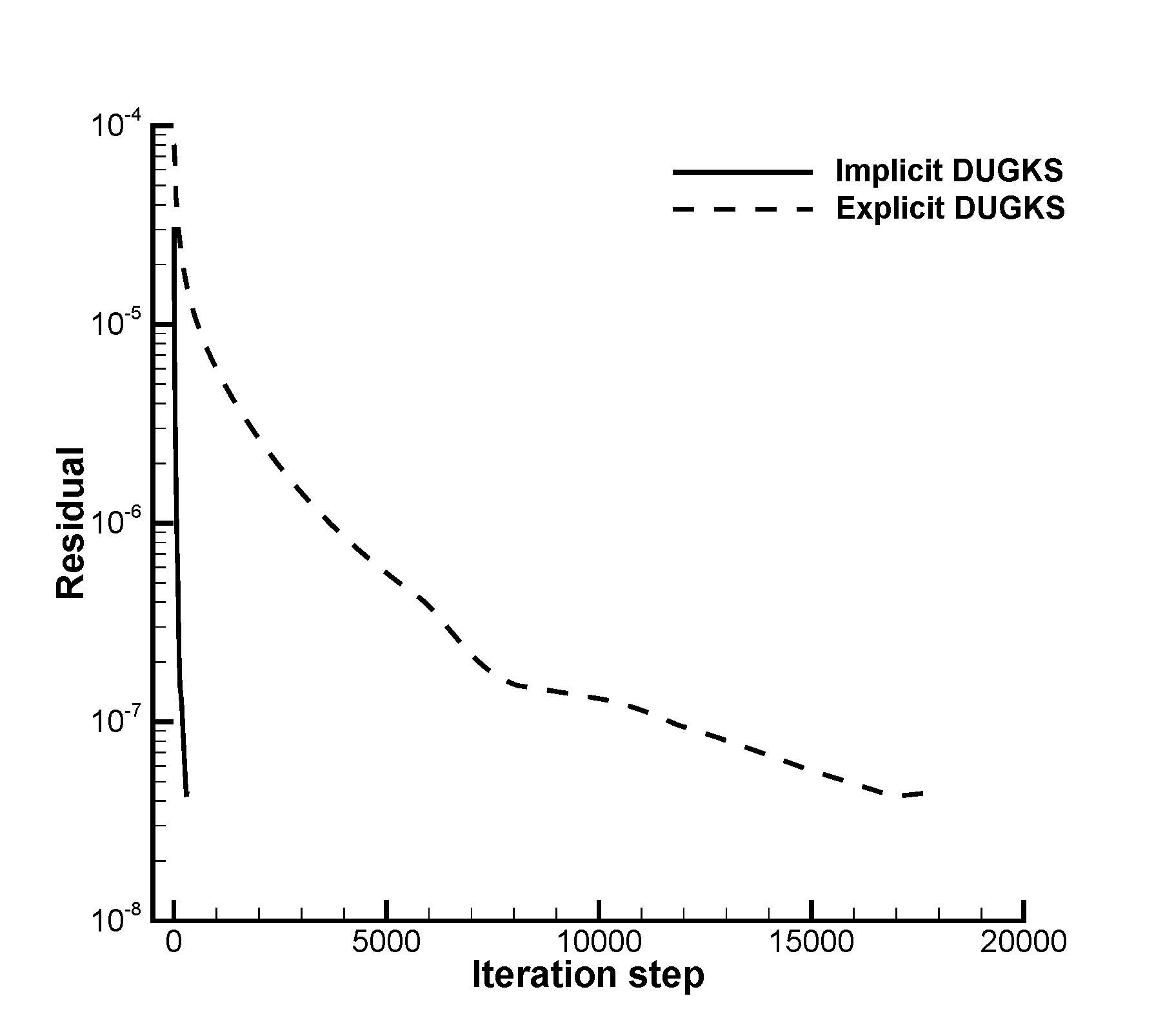}
  \caption{The residual curves of hypersonic circular cylinder flow.}
  \label{fig:Fig21}
\end{figure}

\begin{figure}
  \centering
  \includegraphics[width=0.5\textwidth]{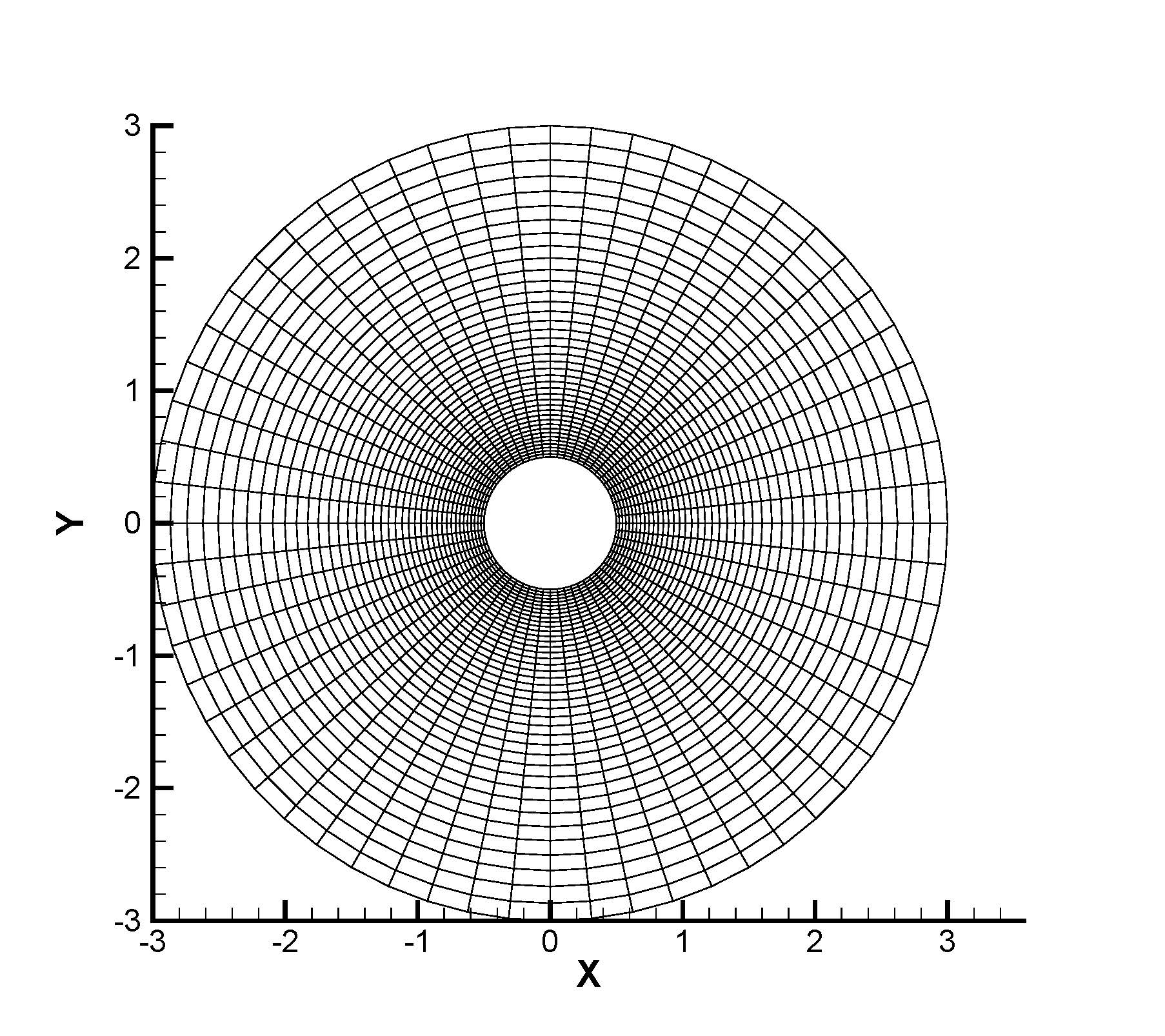}
  \caption{Initial mesh for hypersonic circular cylinder flow.}
  \label{fig:Fig22}
\end{figure}

\begin{figure}
  \centering
  \includegraphics[width=0.5\textwidth]{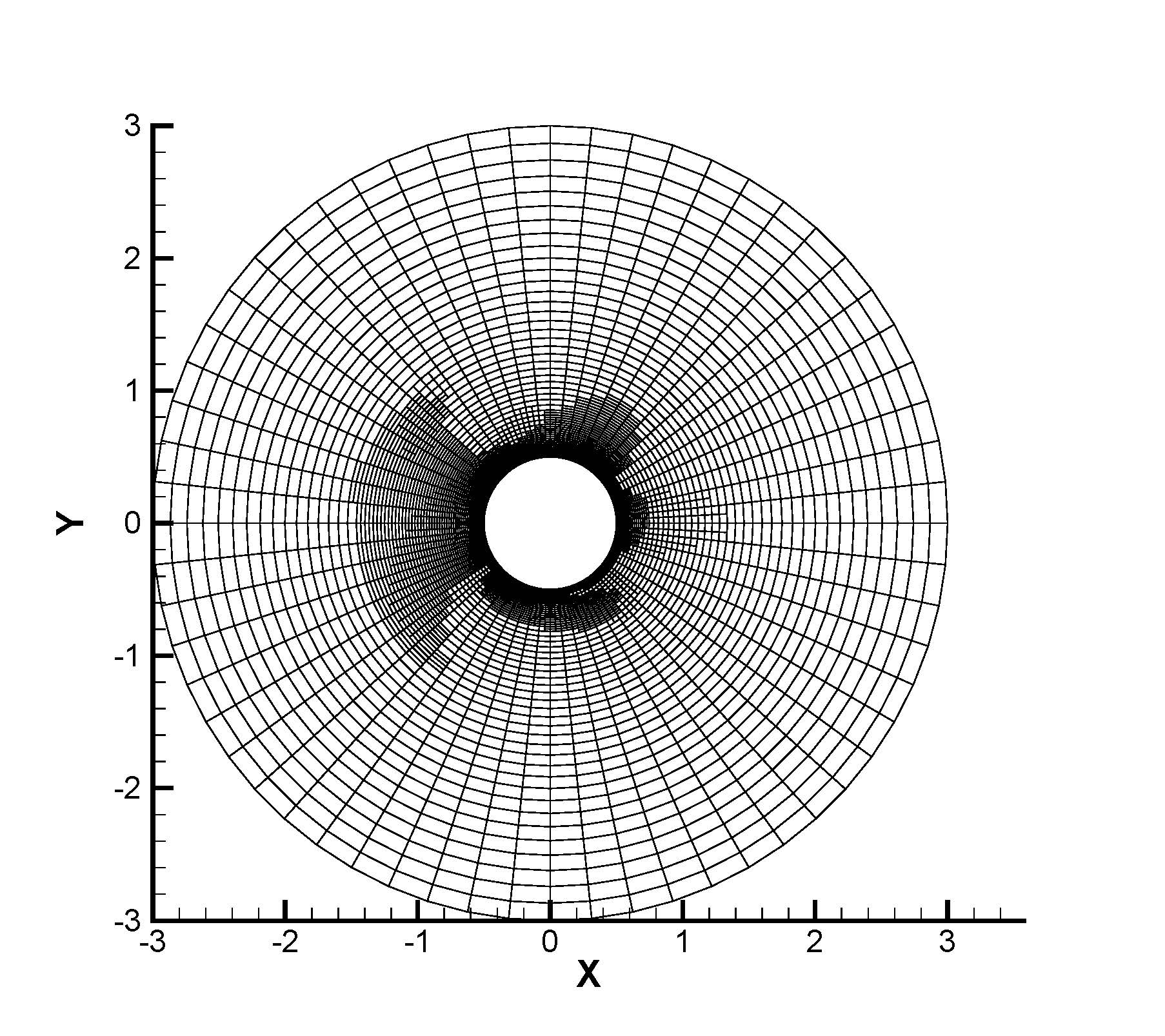}
  \caption{Refined mesh for hypersonic circular cylinder flow. }
  \label{fig:Fig23}
\end{figure}

\begin{figure}
  \centering
  \includegraphics[width=0.5\textwidth]{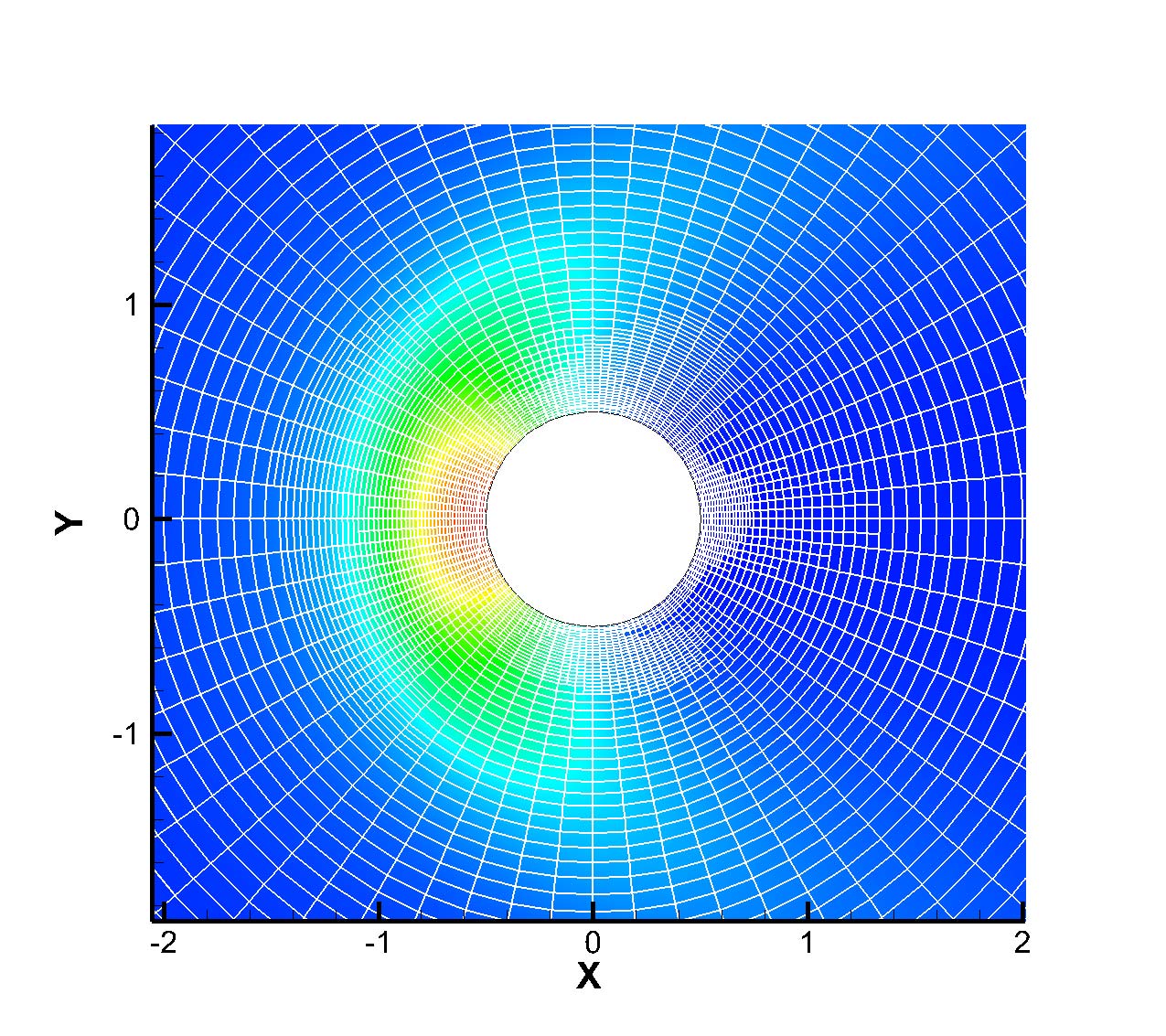}
  \caption{Pressure contour of hypersonic circular cylinder flow under refined mesh.}
  \label{fig:Fig24}
\end{figure}

\begin{figure}
  \centering
  \includegraphics[width=0.5\textwidth]{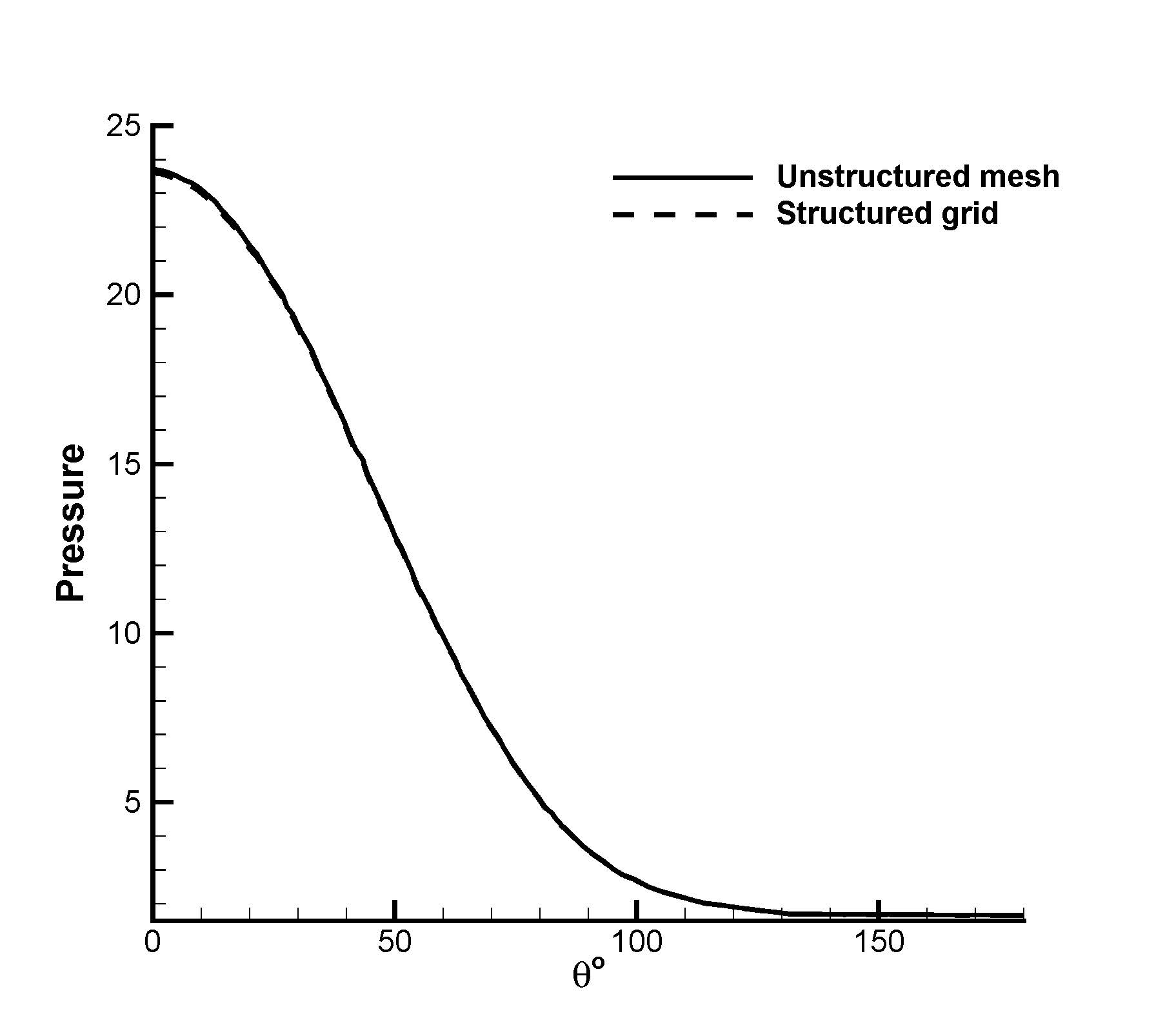}
  \caption{The comparison result for pressure distribution on surface of circular cylinder on different meshes.}
  \label{fig:Fig25}
\end{figure}

\begin{figure}
  \centering
  \includegraphics[width=0.5\textwidth]{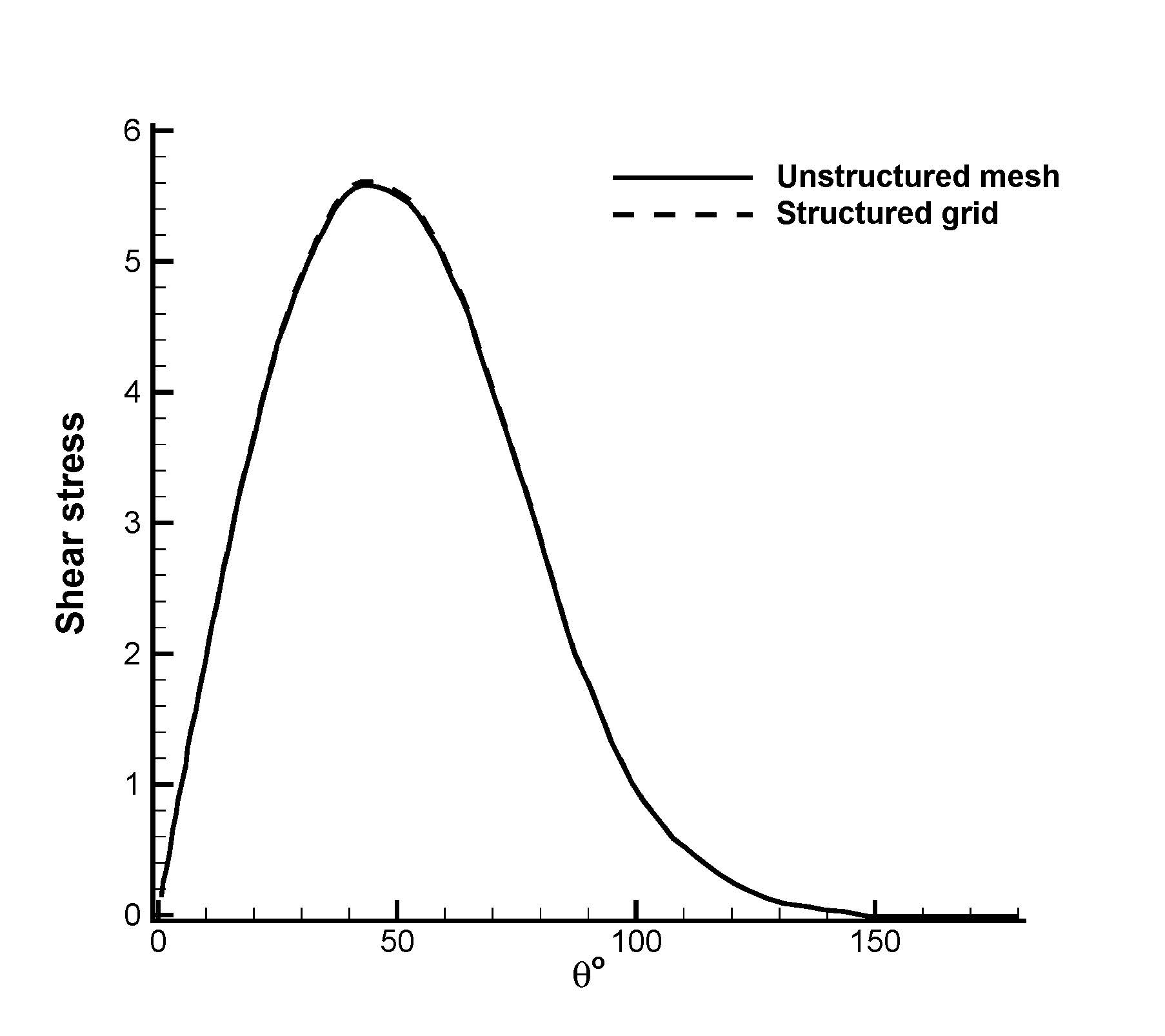}
  \caption{The comparison result for shear stress distribution on surface of circular cylinder on different meshes.}
  \label{fig:Fig26}
\end{figure}

\begin{figure}
  \centering
  \includegraphics[width=0.5\textwidth]{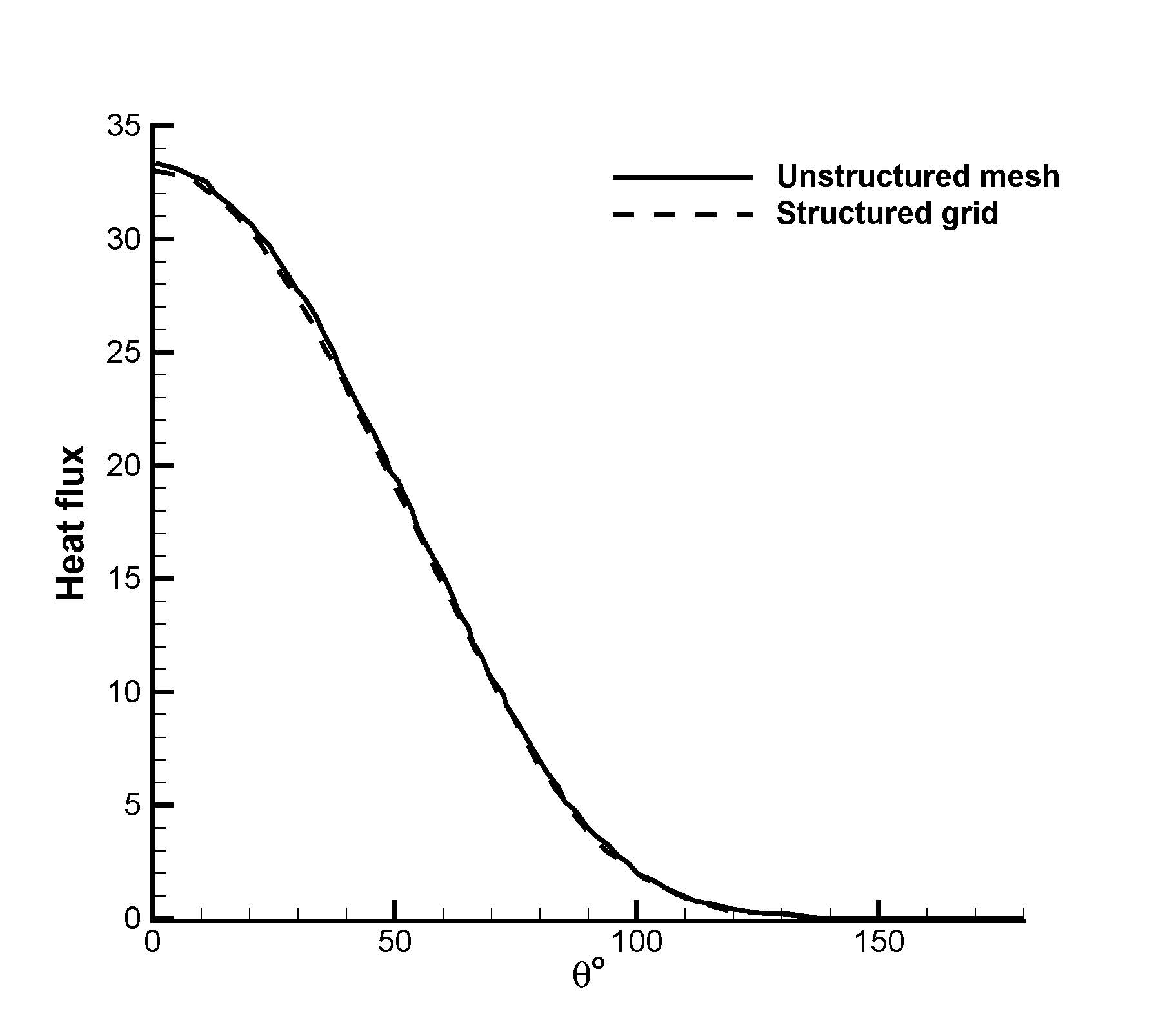}
  \caption{The comparison result for heat flux distribution on surface of circular cylinder on different meshes.}
  \label{fig:Fig27}
\end{figure}

\begin{figure}
  \centering
  \includegraphics[width=0.5\textwidth]{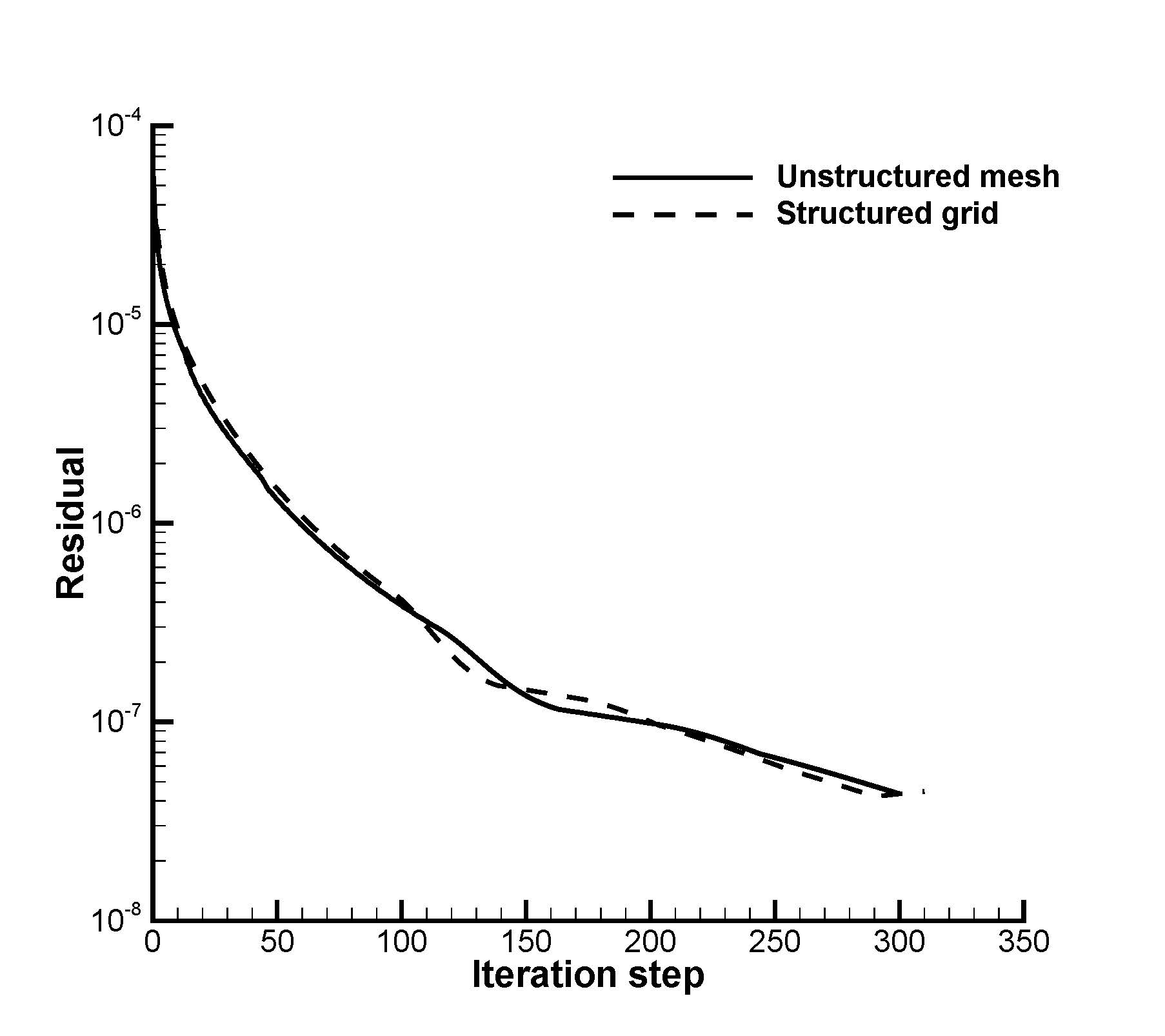}
  \caption{The comparison for residual curves of hypersonic circular cylinder flow on different meshes.}
  \label{fig:Fig28}
\end{figure}

\end{document}